\documentclass[
superscriptaddress,
 amsmath,amssymb,
 aps,
]{revtex4-1}

\usepackage{graphicx}
\usepackage{natbib}
\usepackage{upgreek}
\usepackage{dcolumn}
\usepackage{bm}

\begin{document}

\preprint{APS/123-QED}

\title{Dynamics of a microswimmer near a curved wall: guided and trapped locomotions}

\author{Nima Sharifi-Mood}
 \email{nima.sharifi-mood@siemens.com}
 \affiliation{Siemens PLM Software, Bellevue, WA 98005, USA}
\author{Pablo G. D\'{\i}az-Hyland}%
\affiliation{Department of Chemical Engineering, University of Puerto Rico -- Mayag\"uez, Mayag\"uez, PR 00681, USA}%
\author{Ubaldo M. C\'ordova-Figueroa}
\affiliation{Department of Chemical Engineering, University of Puerto Rico -- Mayag\"uez, Mayag\"uez, PR 00681, USA}%
\begin{abstract}
We propose a combined analytical-numerical strategy to predict the dynamics and trajectory of a microswimmer next to a curved spherical obstacle. The microswimmer is actuated by a slip velocity on its surface and a uniformly valid solution is provided by utilizing the Reynolds reciprocal theorem in conjunction with the exact hydrodynamic solution of translation/rotation of a sphere in an arbitrary direction next to a stationary obstacle. This approach permits the hydrodynamic interaction of the microswimmer and the obstacle to be consistently and accurately calculated in both far and near fields. Based on the analysis, it was shown that while the ``point-singularity solution'' is valid when the microswimmer is far from the obstacle, it fails to predict the correct dynamics when the swimmer is close to the obstacle (i.e. gap size is approximately twice the characteristic length of the microswimmer). Two different paradigms for propulsion, so-called ``squirmer'' and ``phoretic'' models, were examined and for each type of microswimmer, various types of trajectories are highlighted and discussed under which circumstances the swimmer can be hydrodynamically trapped or guided by the obstacle. The analysis indicates that it is always easier to capture a microswimmer in a closed circular orbit with a large sized obstacle ($\sim 20$-$40$ times larger than the microswimmer size) as in this case the magnitude of rotational velocity can be sufficiently large that the swimmer can adjust its distance and orientation vector with the obstacle. 
\end{abstract}
\maketitle
\section{Introduction}
The presence of a boundary modifies the dynamics and trajectories of a large number of microorganisms as well as synthetic self-propelled (active) particles in a counter-intuitive way \cite{Lauga_2006,DiLeonardo25052010,DiLeonardo2011,Volpe,Simmchen_2016,Velegol_2015,Poon_2016}. For example, \emph{E. coli} traces out large circular trajectories in proximity of a solid boundary \cite{Lauga_2006} and its direction of rotation is reversed when it swims close to a stress-free (air-water) interface \cite{DiLeonardo25052010}. Certain types of bacteria tend to accumulate close to solid/liquid (or liquid/liquid) interfaces and traverse along the interface for a prolonged time which in turn has a significant impact in biofilm formation \cite{Kolter_review,Biofilm-rev,Vaccari_2015}. While this behavior has been fully investigated and attributed to hydrodynamic interactions between microswimmers and the boundary \cite{Lauga_2008,spagnolie_lauga_2012}, other factors such as passive interactions (e.g. electrostatics, van der Waals, steric repulsion) and biological aspects (e.g. chemotaxis) can be equally critical. 

There are various models introduced to describe the motion of microorganisms \cite{Lighthill,Najafi,Lauga_Powers}. These abridged models are extremely useful as they are capable of providing insights on the important hydrodynamic features of various classes of swimmers and hence could be readily used to determine their dynamics under confinement or near surfaces. The common theoretical framework used to understand the motion of a microorganism near a planar no-slip wall relies on the singularity (image) solution of the hydrodynamic equations in the creeping flow regime \cite{Lauga_2008,spagnolie_lauga_2012,Spagnolie_Lauga_2015}. Adopting this approach, the leading order contribution to the translational velocity of a swimmer with a force dipolar flow field next to a planar solid wall with a no-slip boundary condition is:
\begin{eqnarray}
{ U_z} =  - \frac{{3p}}{{8{h^2}}}\left( {1 - 3{{\cos }^2}\Xi } \right),
\end{eqnarray}
where $\Xi$ is the incident angle (see the schematic in Fig.~\ref{figure1}), $h$ is the distance from the center of mass of swimmer to the wall and $p$ is the dipole strength ($p>0$ is for a pusher and $p<0$ for a puller) \cite{Lauga_2008,spagnolie_lauga_2012}. The above expression indicates pushers with parallel orientations relative to the wall (i.e. $\Xi \sim {\pi  \mathord{\left/{\vphantom {\pi  2}} \right.\kern-\nulldelimiterspace} 2}$) are hydrodynamically attracted towards the wall. Although the image approach is very intuitive, the solution is reliable only in the far field and it loses its accuracy when the swimmer approaches the wall where higher order flow field modes generated by the microorganism become as significant as the leading order force dipole. 

More recently, advances in nanotechnology and miniaturization enable the design and fabrication of microswimmers capable of taking up energy from their surroundings and converting it into autonomous motion. Propulsion of these particles can be achieved via a number of mechanisms including electrophoresis, thermo- or diffusiophoresis (migration due to temperature and concentration gradients) triggered by onboard processes such as chemical reactions taking place asymmetrically on the surface of the particle \cite{Jiang, Sen2004, Howse2007}. These synthetic microswimmers do compete with their biological counterparts in their ability to swim, transport and operate under non-equilibrium conditions \cite{Schmidt2001,Patra_2013,Wang_2014,Wang_Pul1}. However, in contrast to microorganisms, the locomotion of these particles are driven by an electromagnetic field or spatial variation of the chemical potential of a species in the system that can interact with the surroundings of the particle. This distinctive feature brings another level of complexity in guiding their motion.

Different external mechanisms such as magnetic fields \cite{Burdick_2008,Wang_Magnet}, gravitaxis \cite{Campbell_2013,Bechinger_2014}, and even chemotaxis \cite{Velegol_2007,Baraban_2013} resulting from imposed concentration gradients, have been explored successfully to achieve directed motion in catalytically-driven Janus particles. However, this occurs at the expense of fully automating the particles and at an additional energy cost that could make their application less practical. On the other hand, motivating studies of these particles near surfaces opens the possibility to achieve guided motion without the need of external stimuli \cite{Volpe,Takagi_2014,Simmchen_2016,Velegol_2015,Poon_2016,Wykes_2017}. Volpe \emph{et al.}\cite{Volpe} found experimentally that Au-Si Janus particles undergo self-diffusiophoretic motion upon light illumination when suspended in water/2,6-lutidine solution below the critical temperature as a result of the local asymmetric demixing of the solution. They also found that next to a planar wall, the particle rotates and aligns its orientation vector towards the wall unit normal and propels itself away from the wall. By contrast, Kreuter \emph{et al.} \cite{Kreuter_2013}, Parmar \emph{et al.} \cite{Parmar_2015}, Das \emph{et al.} \cite{Velegol_2015} and Simmchen \emph{et al.} \cite{Simmchen_2016} studied trajectories of spherical polystyrene and Si particles half coated with Pt in hydrogen peroxide and demonstrated that the proximity of a solid planar boundary can quench Brownian rotation of the particles and hence wall-guided motion is achieved. 

There are various theoretical studies addressing self-diffusiophoresis of a particle near a solid planar wall \cite{crowdy_2013,Uspal,Liverpool_2015,Liverpool_2016,Wall_1,Yariv_2016}. First, Crowdy \cite{crowdy_2013} found an exact analytical solution by a conformal mapping technique for the self-diffusiophoresis of a two dimensional Janus disk (with an adsorbing and a producing face for one solute) near a planar wall. It was shown that the translational velocity of the disk can be enhanced next to the wall depending on the particle activity and its orientation. Moreover, the disk could be reoriented and propelled away from the wall for some initial disk orientations. Later on, a more realistic case of spherical Janus particle with constant flux production of a solute near a planar (no-slip) wall was studied numerically utilizing the boundary element method (BEM) \cite{Uspal}. Ibrahim and Liverpool \cite{Liverpool_2015,Liverpool_2016} solved the problem approximately using method of images while Mozaffari \emph{et al.} provided an exact analytical solution adopting bispherical coordinates \cite{Wall_1}. These studies highlighted a couple of important trajectories of a Janus swimmer approaching a planar wall. It was shown that for small reactive particles with small reactive patches $\theta_{\rm{cap}}<90^{\circ}$ on its surface, the swimmer approaches and rotates to reorient until it is reflected from the wall. For larger reactive patches $\theta_{\rm{cap}}>90^{\circ}$, more interesting trajectories were identified as the swimmer can now reach a distance in which its translational velocity in a direction normal to the wall becomes identically zero and its rotation (due to hydrodynamic interaction with the wall) is completely suppressed. This causes the particle to skim parallel to the wall indefinitely which may be of particular interest for guided motions. For even higher reactive patches, the particle loses its mobility completely (translational and rotational velocities in all directions become zero) and it orients axisymmetrically ($\Xi=180^{\circ}$) from the surface and remains in an equilibrium distance indefinitely. It is noteworthy to emphasize that while the particle is at rest in the stationary state, the fluid is not static and hence this particular situation might have an implication in micro-mixing \cite{Uspal}. Moreover, it has been recently shown that the skimming and stationary states are marginally stable for small translational and rotational perturbations of the swimmer \cite{Wall_2}.

Experimental studies of bimetallic rods and Janus spherical swimmers near spherical curved walls reveal they can be captured in orbital paths \cite{Takagi_2014,Poon_2016,Simmchen_2016}. In addition, the time a particle takes to orbit a curved surface has been shown to increase with the fuel concentration (or swimming speed as it is  responsible for faster particle velocity). All of these observations can be originated by the interplay between hydrodynamic interactions, Brownian motion and phoretic effects, in the limit of close proximity of the particle to the surface. 

Takagi \emph{et al.} \cite{Takagi_2014} developed a pure hydrodynamic model based on lubrication analysis to predict the dynamics and trajectory of a Au-Pt rod near a curved spherical wall. The analysis indicates that the presence of a slip velocity on the rod brings about a swimming velocity which is approximately independent of the distance to the obstacle. Furthermore, the lubrication effect was considered as the main mechanism of the attraction of the microswimmers to the wall. By contrast, Spagnolie \emph{et al.} \cite{Spagnolie_Lauga_2015} adopted a far field hydrodynamic model to analyze the dynamics and orbiting of microswimmers next to a large spherical obstacle with no-slip boundary condition. Having utilized the image solution and the F\'axen law, they showed that similar to the planar wall, the presence of the force dipole brings about the attraction of microswimmers next to a curved obstacle. Additionally, it was argued that both ``puller'' and ``pusher'' swimmers can be trapped in an orbital path while the minimum radius of a spherical obstacle to capture a puller is smaller. Recently, Papavassiliou and Alexander \cite{Alexander_2017} developed an exact analytical solution in bispherical coordinates for a squirmer swimmer next to a curved wall. Despite their analysis being solely restricted to axisymmetric cases, the far field solution of Spagnolie \emph{et al.} \cite{Spagnolie_Lauga_2015} was found accurate until the separation distance is of the order of a few swimmer diameters. Nonetheless, inasmuch as orbiting of microswimmers occurs in very small distances to the surface \cite{DiLeonardo2011,Takagi_2014,Poon_2016,Simmchen_2016}, it is important to elucidate the influence of higher order singularities (e.g. source dipole) beside the force dipole term. In addition, there is a need for fundamental understanding of the effect of phoretic interactions between a catalytically-driven particle and inert curved surfaces.

This article provides a model to investigate the dynamics of a spherical microswimmer near a curved stationary obstacle with arbitrary size (or curvature). Regarding locomotion of microorganisms or synthetic active particles, the flow induced by their action is inertialess as the Reynolds number $Re$ is negligible and the hydrodynamics is governed by the Stokes equations \cite{Purcell,Lauga_Powers}. We focus on two general classes of microswimmers namely the ``squirmer'' and the ``phoretic" models where the former is a conventional model used for describing the dynamics of biological microswimmer while the latter is a chemo-mechanical transduction mechanism in which a gradient of solute species in the solution brings about particle propulsion. Regardless of swimmer type, the effect of propulsion can be considered as a ``slip velocity'' at a spherical surface. Decomposing the slip velocity into the suitable basis functions for spherical coordinates and utilizing Reynolds reciprocal theorem (RRT), an exact analytical expression can be deduced for the translational/rotational velocities of the particle next to an obstacle with arbitrary size (curvature). Based on this model, we first discuss the dynamics of both microswimmers and the mechanism adopted for propulsion. Afterwards, we discuss how the near field hydrodynamic interactions with the obstacle can modify the trajectory of the swimmer and subsequently highlight various possible trajectories next to a curved obstacle.  In particular we declare why and under what condition different microswimmers can be captured hydrodynamically in closed orbits. 
\section{Formulation and Solution Techniques} \label{Formulation and Solution Techniques}
\begin{figure}
\centering
\includegraphics[width=0.4 \textwidth]{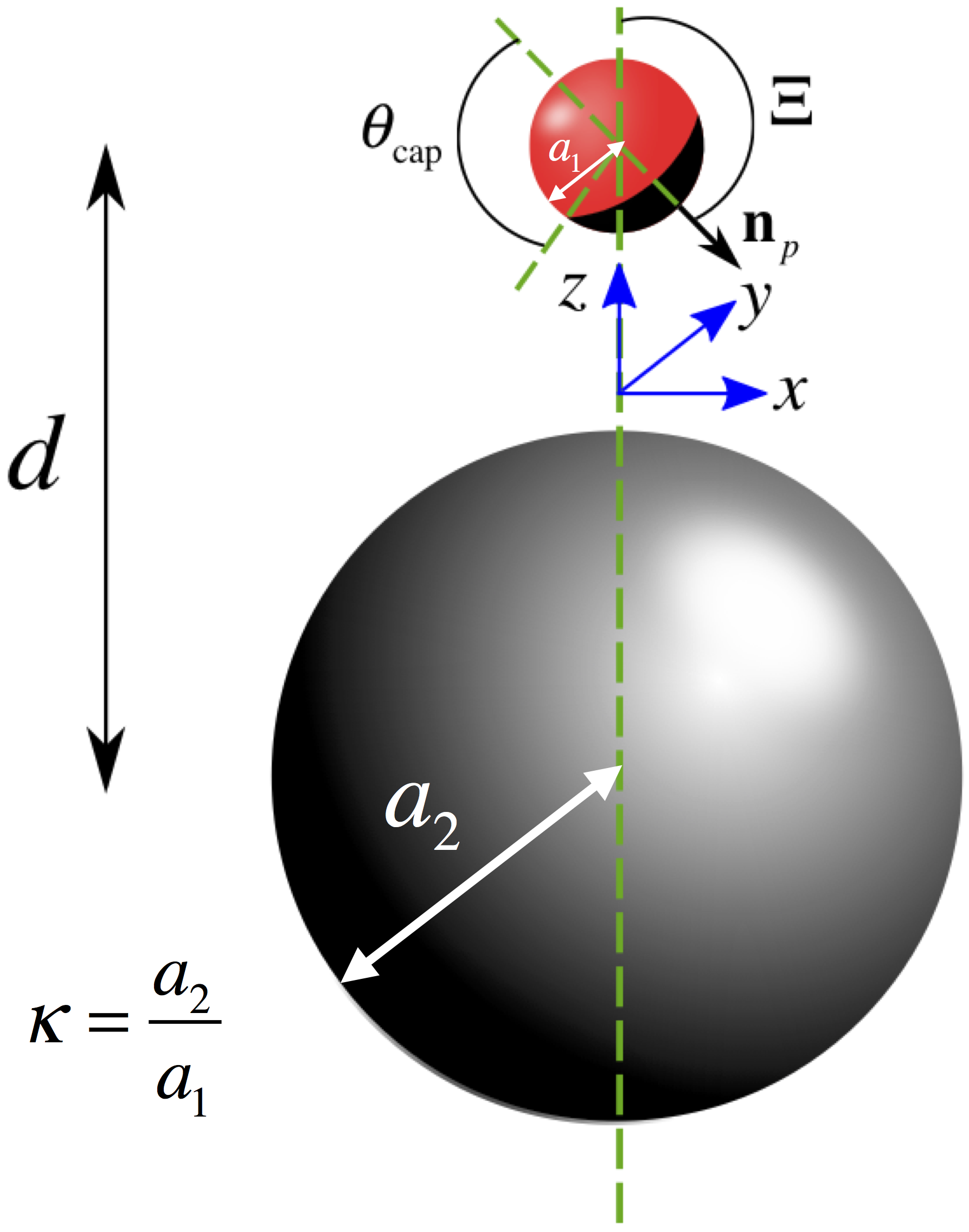}     	    
\caption{\footnotesize{Schematic representation of a spherical swimmer next to a solid spherical curved wall where ${\bf{n}}_p$ is the unit orientation vector of the swimmer, $\Xi$ denotes the relative orientation angle of the particle with respect to the wall and $\theta_{\rm{cap}} $ is an angle determining the active section size of the phoretic swimmer.}} 
\label{figure1}
\end{figure}
We consider a spherical swimmer of radius $a_1$ approaching a solid spherical wall of radius $a_2$ (the size ratio defined as $\kappa  = {{{a_2}} \mathord{\left/{\vphantom {{{a_2}} {{a_1}}}} \right. \kern-\nulldelimiterspace} {{a_1}}}$). The particle swims in an oblique incident with a center-to-center distance of $d=a_1\Delta=a_1(1+\kappa+\delta)$ where $\delta$ is the edge-to-edge separation distance, a relative orientation angle of $\Xi$ and a unit orientation vector ${\bf{n}}_p$ determining the direction of propulsion (see Fig.~\ref{figure1}).  Generally, when a microswimmer traverses next to a boundary, its flow disturbance must satisfy the ``no-slip" boundary condition at the obstacle which in turn creates an induced disturbance which is reflected back to the swimmer and hence modifies the swimmer's dynamics and trajectory. The hydrodynamics of the problem can be formulated irrespective of the propulsion mechanism and hence first the strategy for obtaining the swimming velocity is formulated and discussed.
\subsection{Hydrodynamics}
The dynamics of a spherical swimmer near a curved obstacle can be governed by solving the equations of motion for the swimmer and the surrounding fluid. Considering the external motion of the swimmer, inasmuch as we are dealing with a colloidal sized swimmer, $a_1$, the inertial terms are negligible ($Re={{\rho U_ca_1} \mathord{\left/{\vphantom {{\rho U_ca_1} \mu }} \right.\kern-\nulldelimiterspace} \mu }\ll1$ where $U_c$ is the  velocity of isolated microswimmer) and hence the hydrodynamic equations in an incompressible, Newtonian fluid can be written as
\begin{eqnarray}
&{\bm{\nabla}}  \cdot {\bf{v}} = 0,\\
&-{\bm{\nabla}}  p + {{\bm{\nabla}}  ^2}{\bf{v}} = 0,
\end{eqnarray}  
where $p$ (scaled with $\sim{{\mu {U_c}} \mathord{\left/{\vphantom {{\mu {U_c}} a_1}} \right.\kern-\nulldelimiterspace} a_1}$) and $\bf{v}$ (scaled with $\sim U_c$) account for non-dimensional pressure and velocity vector respectively. The above equations are subjected to the following boundary conditions to yield a well-posed problem
\begin{eqnarray}
&{\bf{v}}({\bf{x}} \in S_c) = {\bf{v}}_s + {{\bf{U}}} + {{{\bm{\Omega} }}} \times ({\bf{x}} - {{\bf{x}}_0}),\\
&{\bf{v}}({\bf{x}} \in S_w) = {\bf{0}},\\
&{\bf{v}}(\left| {\bf{x}} \right| \to \infty ) \sim {\bf{0}},
\end{eqnarray}
where $\bf{x}$ is the position vector in the lab frame, $\bf{x}_0$ represents the swimmer center of mass, $S_c$ and $S_w$ denote the surface domains of the swimmer and the wall respectively, ${\bf{v}}_s$ is the slip velocity depending on the propulsion mechanism of the swimmer and ${{\bf{U}}}$ and ${\bm{\Omega }}$ are the terminal translational and rotational velocities yet to be determined. These values can be obtained by invoking the fact that the swimmer is force and torque free, i.e. the total force ${\bf{F}}_T$ and torque ${\bf{T}}_T$ on the swimmer is zero
\begin{align}
&{\bf{F}}_T= \oint_{{S_c}} { {{\bf{e}}_n \cdot {\bm{\sigma}}~ dS}}  = 0,\\
&{\bf{T}}_T = \oint_{{S_c}} {{({\bf{x}} - {{\bf{x}}_0}) \times {\bf{e}}_n \cdot {\bm{\sigma}}~ dS}}  = 0,
\end{align}
where $\bm{\sigma}$ is the total hydrodynamic stress tensor, ${\bf{e}}_n$ is the unit normal vector pointing outward from the swimmer surface and $dS$ denotes the area element. Notice that the obstacle has been presumed to be stationary with no-slip boundary condition. Additionally, considering a deterministic motion where the effect of Brownian motion of the swimmer is negligible, the hydrodynamic equations along with the boundary conditions are all linear and hence can be solved analytically utilizing the standard technique developed for Stokes flow in bispherical coordinates \cite{JFM2,Wall_1}. Here, we take advantage of an alternative approach which relies on utilizing the RRT \cite{teubner,Stone_Samuel,Masood,Elfring_RRT} to calculate the  translational and rotational swimming velocities as it was shown that these values can be evaluated as
\begin{eqnarray}
{\bf{\cal{U}}} = {{\bf{\cal{M}}}}\cdot\oint_{{S_c} \cup {S_w}} {{{\bf{v}}_s} \cdot ({{\bf{e}}_n} \cdot {\bf{\cal{T}}})dS} ,\label{compact}
\end{eqnarray}
where $\bf{\cal{U}}$ is a column vector $[6\times 1]$ that contains the translational and rotational velocities of the swimmer, $\bf{\cal{M}}$ is the hydrodynamic mobility symmetric positive-definite matrix, a $[6\times 6]$ matrix including the exact hydrodynamic interaction of a spherical particle near a curved wall, and ${\cal{T}}$ denotes the triadic hydrodynamic stress fields (a $[3 \times 3] \times 6$ matrix) associated with translations and rotations of a spherical particle next to a curved wall in Cartesian coordinates.

Having determined the translational and rotational velocities of the swimmer, the trajectory of the swimmer can be obtained by integrating the following equations
\begin{align}
&\frac{{d{\bf{x}}_0}}{{dt}} = {\bf{U}},\label{traj-u}\\
&\frac{{d{{\bf{n}}_p}}}{{dt}} = {\bm{\Omega} }  \times {{\bf{n}}_p},\label{traj-o}
\end{align}
where $t$ accounts for non-dimensional time (scaled with $\sim{{{a_1}} \mathord{\left/{\vphantom {{{a_1}} U_c}} \right.\kern-\nulldelimiterspace} U_c}$). The above equations can be integrated with an explicit numerical scheme to yield the swimmer trajectory. 
As noted earlier, the slip velocity on the swimmer surface ${\bf{v}}_s$ depends on the propulsion mechanism of the swimmer. Below we discuss each one of them separately.
\subsection {Squirmer Swimmer}
Squirmer model is a generic model used to represent microorganism motion \cite{Lighthill,Blake_1971,Lauga_Pak_2014}. The model was originally inspired by the motion of ciliated organism so-called ``opalina'' which traverses fluids by beating cilia. In this model, a ``slip velocity'' represents the action of beating cilia on a sphere encompassing them and it can be systematically cast into appropriate basis functions according to 
\begin{eqnarray}
{{\bf{v}}_s} = {{{\bf{n}}_p} \cdot \left( {{{\bf{e}}_{\bf{x}}}{{\bf{e}}_{\bf{x}}} - {\bf{I}}} \right)}\sum\limits_{n = 1}^\infty  {\frac{2}{{n(n + 1)}}{B_n}{\partial_{\theta}P_n}({{\bf{n}}_p} \cdot {{\bf{e}}_{\bf{x}}})},
\end{eqnarray}
where ${{\bf{e}}_{\bf{x}}}$ is the unit position vector, ${\bf{I}}$ is the identity tensor, $B_n$ is the $n$th mode of the tangential slip velocity, $\partial_{\theta}P_n$ represents the derivative of the Legendre polynomial and $\theta$ is the polar angle measured from the orientation vector of the swimmer. In Appendix A (Sec.~\ref{appendix}), the detailed flow field around a single squirmer model is recapitulated \cite{Blake_1971,Pedley}. Here the focus is solely on the first two squirming modes and hence the magnitude of the tangential velocity can be written as
\begin{eqnarray}
{{\bf{v}}_s}(\theta ) = ({B_1}\sin \theta  + {B_2}\sin \theta \cos \theta){\bf{e}}_{\theta},\label{squirmer-slip}
\end{eqnarray}
where $B_1$ and $B_2$ are the first two non-zero modes. The terminal velocity of the squirmer model is governed by $B_1$ (${\bf{U}}={2 \mathord{\left/{\vphantom {2 3}} \right.\kern-\nulldelimiterspace} 3}{B_1}{{\bf{n}}_p}$) whereas $B_2$ does not have any influence on the magnitude of velocity but it determines the magnitude of stress (known as stresslet) as a result of propulsion and hence it only affects the flow field around the swimmer (see Appendix A for details). Although both of these coefficients can be intrinsically time-dependent, constant values of these modes were considered. The ratio of the two modes $\gamma = {{{B_2}} \mathord{\left/{\vphantom {{{B_2}} {{B_1}}}} \right.\kern-\nulldelimiterspace} {{B_1}}}$ determines different classes of squirmers: $(1)$ $\gamma > 0$ is a puller generating thrust in front of the body such as \emph{Chlamydomonas}; $(2)$ $\gamma < 0$ corresponds to a pusher generating thrust behind the body such as most bacteria; $(3)$ $\gamma = 0$ corresponds to a neutral squirmer, generating a symmetric potential flow field, such as \emph{Volvox} and $(4)$ $\gamma \to \infty$ corresponds to a squirmer which cannot swim but it traverses the surrounding fluid by creating a flow field due to its stresslet. This swimmer model is often called ``shakers'' and has been used in the literature to understand the collective motion of microorganisms and their motions near boundaries \cite{Llopis_2008,Lauga_2008,Spagnolie_Lauga_2015}. The flow field induced by a squirmer model up to the leading order is a force dipole and attenuates as $\sim r^{-2}$ except a neutral squirmer for which the flow field can be represented with a (potential flow) source dipole and hence the velocity field decays as $\sim r^{-3}$ where $r$ is the radial distance from center of mass of the swimmer.

\subsection{Diffusiophoretic Swimmer}
The propulsion for this class of swimmer is achieved by the asymmetric distribution of interactive solutes around the particle. In this case, the swimmer itself creates and sustains the concentration gradient of a neutral solute by a chemical reaction at its reactive surface. The dimensional slip velocity has been shown to be provided according to \cite{Howse2007,Nima_1,Lauga_Peclet} 
\begin{eqnarray}
{{\bf{\tilde v}}_s} =   b\left( {{\bf{I}} - {{\bf{e}}_n}{{\bf{e}}_n}} \right) \cdot {\bm{\nabla}} \tilde n,
\end{eqnarray}
where $\tilde n$ is the dimensional solute distribution and $b$ is the slip coefficient defined as\cite{Howse2007}
\begin{eqnarray}
b = \frac{{{k_B}T{L^2}}}{\mu }\int_0^\infty  {\left[ {\exp ( - \frac{\psi }{{{k_B}T}}) - 1} \right]dy'},
\end{eqnarray}
where $k_B$ is the Boltzmann constant, $T$ is the absolute fluid temperature, $L$ is the interaction layer, $y'$ is the distance to the particle edge and $\psi$ is the net potential interaction of the solute molecules with the particle in the presence of solvent. For a hard-sphere potential between solutes and swimmer, this reduces to $b =  - {{{k_B}T{L^2}} \mathord{\left/{\vphantom {{{k_B}T{L^2}} {2\mu }}} \right.\kern-\nulldelimiterspace} {2\mu }}$. 

The slip velocity depends on solute distribution around the swimmer and hence to obtain the swimmer velocity, the solute distribution must be evaluated first. Here, a simple form for the activity of the swimmer is adopted which presumes a uniform flux production $j_0$ of one solute on the active section of the swimmer. Furthermore, the solute distribution is governed by solving the (non-dimensional) mass balance conservation in the Lagrangian frame for the solute according to
\begin{eqnarray}
Pe_s\left( {\frac{{\partial n}}{{\partial t}} + {\bf{v}} \cdot {\bm{\nabla}} n} \right) = {{\bm{\nabla}} ^2}n,
\end{eqnarray}
where $n = {{\tilde nD} \mathord{\left/{\vphantom {{\tilde nD} {{j_0}{a_1}}}} \right.\kern-\nulldelimiterspace} {{j_0}{a_1}}}$, $D$ is the solute diffusivity and $Pe_s$ is the solute P\'eclet number, $Pe_s = {{a_1{U_c}} \mathord{\left/{\vphantom {{a{U_c}} D}} \right.\kern-\nulldelimiterspace} D}$ identifying the relative importance of the time scale for solute advection (${a_1 \mathord{\left/
 {\vphantom {a {{U_c}}}} \right.\kern-\nulldelimiterspace} {{U_c}}}$) to the time scale for diffusion of product solute (${{{a_1^2}} \mathord{\left/{\vphantom {{{a_1^2}} D}} \right.\kern-\nulldelimiterspace} D}$). 

The above equation can be further simplified to a diffusion equation provided that $Pe_s \ll 1$, i.e.
\begin{eqnarray}
{{\bm{\nabla}} ^2}n = 0,\label{diffusion}
\end{eqnarray}
with boundary conditions defined as
\begin{align}
{{\bf{e}}_n} \cdot {\bm{\nabla}} n({\bf{x}} \in {S_c}) &= u({\bf{x}}),\label{bc1-diffusion}\\
n({\bf{x}} \in {S_w}) &= 0,\label{bc2-diffusion}\\
n(\left| {\bf{x}} \right| \to \infty ) &\sim 0,\label{bc3-diffusion}
\end{align} 
where $u({\bf{x}})$ is the coverage function which is equal to unity on the active section of the swimmer and zero elsewhere. The second boundary condition states that the wall is impenetrable to the solute and finally by the third boundary condition, we assume the solute concentration in a region far from the swimmer is zero. 

Eq.~\ref{diffusion} with boundary conditions (Eqs.~\ref{bc1-diffusion}$-$\ref{bc3-diffusion}) is a linear, elliptic PDE and the solution can be found analytically providing a judicious choice of a curvilinear coordinate. In this case, the general three-dimensional solution can be found in a conventional bispherical curvilinear coordinate ($\alpha$, $\beta$, $\phi$). Constant values of $\beta$ define surfaces of the spherical swimmer ($\beta = \beta_1>0$) and the curved spherical wall ($\beta=\beta_2<0$) and the fluid domain is constrained to $0 \le  \alpha  \le \pi $, $\beta_2< \beta < \beta_1 $ and the azimuthal angle of $0\le \phi  \le2\pi $ can be represented as constant values of $\beta$. Thus the solution to the Laplace equation can be given as \cite{Nima_3}
\begin{align}
&n(\alpha ,\beta ,\phi ) = \sqrt {\cosh \beta  - \cos \alpha }\nonumber\\
 &\sum\limits_{m = 0}^\infty  {\sum\limits_{n = m}^\infty  {\left[ {({{\tilde A}_{nm}}\cosh (n + 0.5)\beta  + {{\tilde B}_{nm}}\sinh (n + 0.5)\beta )\cos m\phi  +} \right.}}\nonumber \\
&\left. {({{\tilde C}_{nm}}\cosh (n + 0.5)\beta  + {{\tilde D}_{nm}}\sinh (n + 0.5)\beta )\sin m\phi } \right]P_n^m(\cos \alpha ),\label{laplacian}
\end{align}
where ${{\tilde A}_{nm}}$, ${{\tilde B}_{nm}}$, ${{\tilde C}_{nm}}$ and ${{\tilde D}_{nm}}$ are unknown coefficients which can be evaluated by imposing the boundary conditions on the swimmer and the curved wall, and $P_n^m(\cos\alpha)$ is the associated Legendre function of the first kind. Having utilized the orthogonality of trigonometric functions and the associated Legendre polynomials, we have a set of recursive relationships which must be solved numerically. These recursive relations were given elsewhere \cite{Nima_3}.
\subsection{Numerical Solution}
The exact hydrodynamic analysis of two spherical particles with arbitrary radii in the Stokes flow regime relies on decomposing the arbitrary motions of two particles into a dozen of fundamental sub-problems. These problems describe translations and rotations of the particles in the parallel and perpendicular directions relative to the center-to-center line of particles \cite{hapbren83,Kim}. The solution to all of these sub-problems (except translational and rotational motions along and around the center-to-center respectively \cite{Jeffery,Spielman}) can be expressed in terms of four double infinite sums. Each one of the sums has two sets of unknown coefficients that needed to be determined based on the boundary conditions imposed on particle surfaces \cite{Majumdar}. To obtain the unknown coefficients of these infinite sums and consequently the detailed flow field (e.g. velocity profile, stress profile, etc) as well as the components of hydrodynamic mobility (or resistance) tensor for each one of the sub-problems (translations and rotations of the spheres parallel to the center-to-center lines and around the third axis respectively), we must solve a linear system ($\bf{A}{x}=\bf{B}$) which consists of $8$ sets of recursive formula for $8$ sets of unknowns. We optimized the numerical calculation by truncating these series according to the fact that all coefficients vanish for sufficiently large number of terms. Additionally, more terms in the truncated series were kept for small separation distances to reach numerically accurate and stable results. The series solution however converges slowly for very small separation distances which results in a large condition number for the matrix $\bf{A}$ and consequently deteriorate achieving a solution for the linear system. Similar difficulty occurs in determining the concentration field and hence to avoid numerical instability, all of our trajectories were calculated for non-dimensional separations greater than $0.01$. Below this distance, it is assumed the swimmer is in contact with the obstacle (notice for diffusiophoretic swimmer, the slip velocity argument assumption may no longer be valid for very small gaps). Our numerical results were fully validated against the tables and figures given in a number of studies in the literature and our previous works \cite{Spielman,Majumdar,Wall_1,Nima_3}. 

All variables in the integral given in Eq.~\ref{compact} were calculated based on analytical relations, however the integrals were computed numerically based on Simpson's rule. Furthermore, the trajectory of the swimmers were determined by numerical integration of Eqs.~\ref{traj-u} and \ref{traj-o} through implementation of an explicit four-step Adams-Bashforth algorithm with a reasonably small time step.

\section{Results and Discussions}
The trajectory of the swimmer is a function of its initial orientation vector and distance with respect to the obstacle. Here, we choose the initial orientation vector of the particle to be in the $x - z$ plane and inasmuch as we are dealing with axisymmetric swimmers in this study, the swimmer stays in the $x - z$ plane indefinitely. 

\subsection{Passive particle with a constant external force along its orientation vector}
\begin{figure}
\centering
\includegraphics[width=0.7\textwidth]{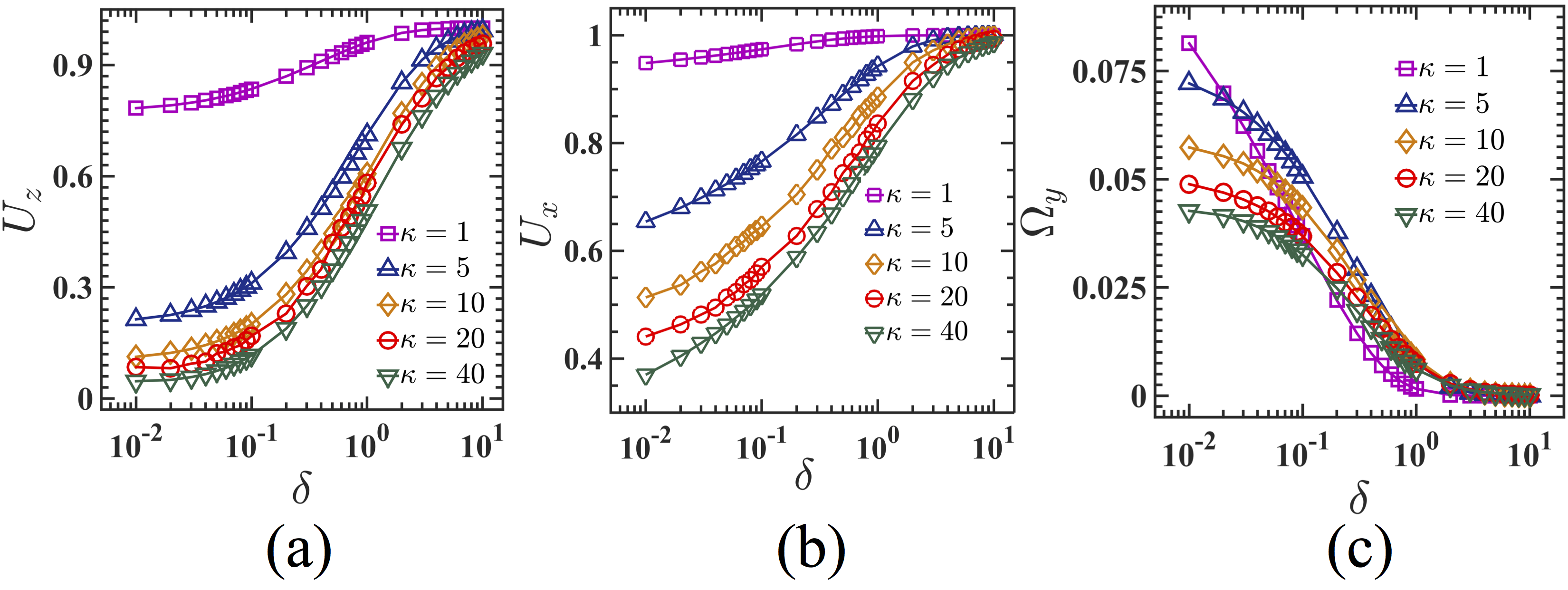}     	    
\caption{\footnotesize{The non-dimensional velocities of a passive particle under a constant force along its orientation vector as a function of non-dimensional edge-to-edge separation distance $\delta$ next to a curved wall (with various degrees of curvature) for inclination angles of $(a)$ $\Xi = 180^{\circ}$ (or $\Xi = 0^{\circ}$), $(b)$ $\Xi = 90^{\circ}$ and $(c)$ $\Xi = 90^{\circ}$}.} 
\label{figure_2}
\end{figure}
To obtain a better insight on the role of hydrodynamic drag on the dynamics of a spherical particle near a curved obstacle, we first consider the motion of a passive particle near a stationary obstacle under a constant external force along its orientation vector. Here, the magnitude of the force is independent of swimmer distance (or its relative orientation) with the obstacle. In this case, the velocity field in a region far away from the particle is a stokeslet plus a source dipole decaying as $\sim r^{-1}$. Additionally, since there is no external torque exerted on the particle, particle rotation does occur only in the vicinity of the obstacle due to hydrodynamic interactions with the no-slip boundary. For axisymmetric incidents ($\Xi=0^{\circ}$ and $\Xi=180^{\circ}$), the passive particle does not rotate and the magnitude of its translational velocity is independent of the tilt angles; i.e. due to the reversibility of Stokes equation \cite{hapbren83,Purcell}, the speed in a case where the particle approaches the wall ($\Xi=180^{\circ}$) is exactly identical to the situation where it moves away from the wall ($\Xi=0^{\circ}$). Fig.~\ref{figure_2}(a) shows the non-dimensional translational velocity (in a direction parallel to the center-to-center line, $z$ direction) of a passive particle under a constant force along its orientation vector for various curvatures of the obstacle. The translational velocity in all cases weakens as the separation distance becomes smaller which is a direct consequence of increase in the viscous drag resistance. For an oblique incident, the asymmetric hydrodynamic force with respect to the swimmer's center of mass brings about a rotation for the swimmer. Fig.~\ref{figure_2}(b) and (c) exhibit the non-dimensional translational and rotational velocites of the particle as a function of its separation distance for a typical tilt angle of $90^{\circ}$ respectively. It can be seen from Fig.~\ref{figure_2}(b) that the translational velocity behaves similarly as in the axisymmetric case; i.e. the speed becomes smaller as it approaches the wall, however, notice the rate at which the translational speed decreases is weaker (in all cases) relative to the axisymmetric case (compare Fig.~\ref{figure_2}(a) and Fig.~\ref{figure_2}(b)) and hence translation in a direction parallel to the wall is always much easier relative to the translation in a direction perpendicular to the wall as the hydrodynamic drag resistance, for small separations ($\delta \ll 1$) in the latter, scales as $\sim {\delta ^{ - 1}}$ while in the former, it scales $\sim - \log (\delta )$. Furthermore, the hydrodynamic resistance is much stronger for a less curved (flat) wall and hence the particle can translate much faster next to a smaller obstacle (smaller $\kappa$). On the other hand, the particle rotates faster when it translates closer to the wall (see Fig.~\ref{figure_2}(c)) as in this case the hydrodynamic drag force becomes more asymmetric with respect to the swimmer's center of mass. The speed of rotation is smaller for a flatter wall (larger values of $\kappa$) nonetheless regardless of the curvature of the obstacle, the rotational velocity is always positive; i.e. the particle always rotates in a way that it tends to align its orientation vector perpendicular to the wall ($\Xi=180^{\circ}$). This is an important feature of this particular model as the presence of the wall does not allow the particle to be scattered.  

\subsection{Squirmer swimmer}
\begin{figure}
\centering
\includegraphics[width=0.7 \textwidth]{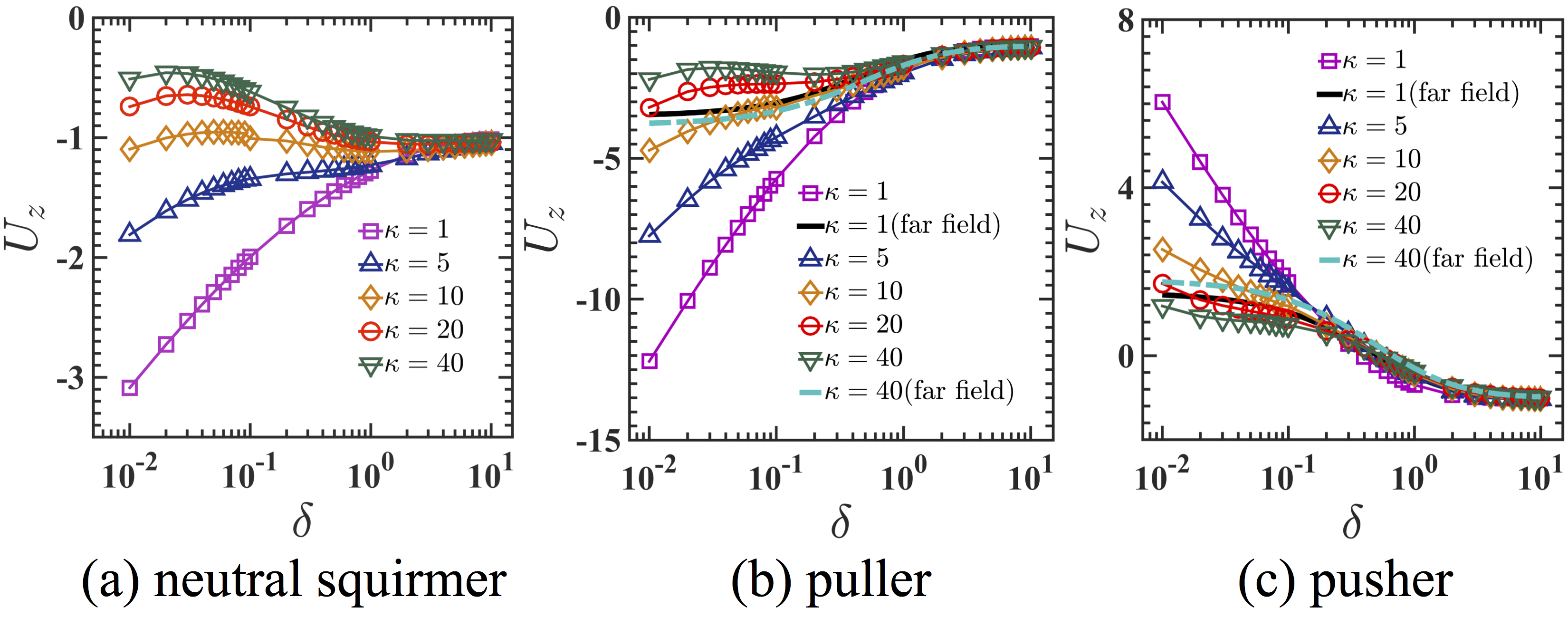}     	    
\caption{\footnotesize{The non-dimensional velocities of a squirmer model as a function of non-dimensional edge-to-edge separation distance $\delta$ approaching a curved wall (with various degrees of curvature) with the tilt angle of $\Xi = 180^{\circ}$ for $(a)$ a neutral squirmer, $(b)$ a pusher ($\gamma=-5$) and $(c)$ a puller ($\gamma=5$).}} 
\label{figure_3}
\end{figure}
The dynamics of a squirmer next to a curved obstacle is complex and the behaviors of the puller, pusher and neutral squirmer next to an obstacle are somewhat different. Fig.~\ref{figure_3} demonstrates the non-dimensional velocity of a squirmer next to a curved wall as a function of the separation distance for an axisymmetric motion where $\Xi = 180^{\circ}$. We can immediately perceive that the squirmer can swim faster (see Fig.~\ref{figure_3}(a) and (b)) or even reverse its swimming direction (Fig.~\ref{figure_3}(c)) as it approaches the obstacle. While the neutral squirmer approaches ($\Xi=180^{\circ}$) and moves away ($\Xi=0^{\circ}$) from the obstacle with identical speeds due to its symmetric flow field (see Fig.~\ref{figure_1_appendix}(b)), the magnitude of axisymmetric propulsive velocity for the puller and pusher not only depends on the separation distance but also on the orientation of the swimmer's head with respect to the wall ($\Xi=0^{\circ}$ or $\Xi=180^{\circ}$). A puller increases its speed as it approach the wall and by contrast a pusher (with an identical strength) reduces speed as it approaches the obstacle and beyond a separation distance it reverses its direction of motion and tends to be repelled from the wall, this particular trend compels the puller to collide with the wall and the pusher to gain a stationary state at an equilibrium distance with respect to the wall. These distinctive behaviors for the puller and pusher rely on the flow field around these two classes of squirmers that are completely different: while the region of high stress for a puller is at swimmer's head (front), the region of high stress is at its tail (back) for a pusher (see Fig.~\ref{figure_1_appendix}(a) and (c) in the Appendix A). When a puller (pusher) approaches a wall with an inclination angle of $\Xi=180^{\circ}$ (i.e. from its front side), it exposes its high (low) region of stress to the wall. The presence of a no-slip wall always enhances the magnitude of stress locally in the gap region and hence the propulsive force would be increased (reduced) for a puller (pusher) as it approaches the wall. This effect is so significant for the pusher in very small gap sizes that it ultimately forces the swimmer to reverse its direction of motion and the swimmer moves away from the wall.  The axisymmetric motion of a neutral squirmer is similar to the puller (see Fig.~\ref{figure_3}(a) and (b)) however the influence of the wall is significant only at small distances as the flow field generated by the neutral squirmer is a source dipole which attenuates faster $\sim r^{-3}$ and therefore the role of boundary comes into play only when the swimmer is sufficiently close. In all cases, a squirmer swims slower next to a larger obstacle (greater $\kappa$) at small separations as the hydrodynamic drag resistance (lubrication force) is greater than the resistance of a smaller obstacle. The far field solution $\kappa=1$ and $\kappa=40$ ( Eq.~39 and 40 in appendix B of the paper by Spagnolie \emph{et al.} \cite{Spagnolie_Lauga_2015}) was also plotted in Fig.~\ref{figure_3}(b) and (c). While the far field solution is accurate up to a distance equivalent to the swimmer diameter, it fails to accurately capture both puller and pusher swimming velocities at smaller distances even qualitatively. At small distances, the far field solution predicts very little dependency on the curvature of the obstacle. It also demonstrates the motion close to the larger obstacle ($\kappa=40$) is faster relative to the smaller one ($\kappa=1$) which is not correct. Additionally, the far field solution for a neutral squirmer would be a constant value regardless of the distance and the obstacle size while the results shown in Fig.~\ref{figure_3}(a) exhibit a strong dependency on both. This inaccuracy originates from the presence of higher modes considered in the squirmer model. The squirmer model has two additional higher order modes, attenuating as $\sim r^{-3}$ and $r^{-4}$, beside the force dipolar term due to its finite size (see Eqs.~\ref{vr_sq} and \ref{vt_sq} in the Appendix A). These terms are negligible when the swimmer is far from the boundaries, however they become as important as the force dipole term close to the boundary and hence modify the swimmer's dynamics.

\begin{figure}
\centering
\includegraphics[width=0.65 \textwidth]{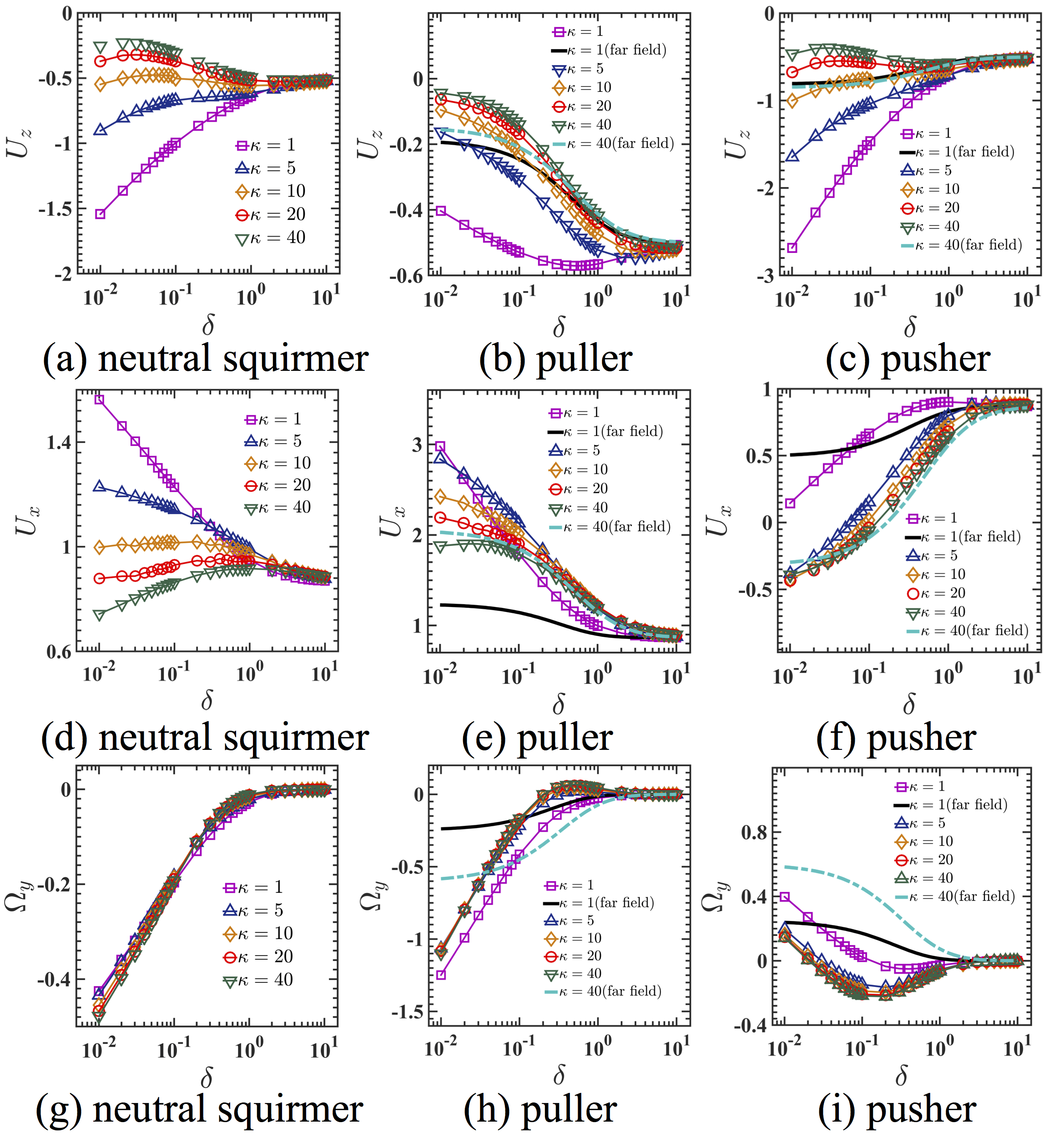}     	    
\caption{\footnotesize{The non-dimensional velocities of a squirmer model as a function of non-dimensional edge-to-edge separation distance $\delta$ next to a curved wall (with various degrees of  curvature) for the inclination angle of $\Xi = 120^{\circ}$. The puller and pusher strengths are $\gamma=5$ and $\gamma=-5$.}} 
\label{figure_4}
\end{figure}
Fig.~\ref{figure_4} demonstrates the non-dimensional translational and rotational velocities of the squirmer model for an asymmetric orientation with respect to the wall ($\Xi=120^{\circ}$). In this case, the swimmer orientation vector is no longer aligned with the wall normal vector. Regarding the translational velocity in the normal direction to the wall $U_z$, a pusher swims faster as it approaches the wall while a puller reduces its speed (see Fig.~\ref{figure_4} (a)-(c)). The swimming velocities in the $x$ direction are shown in Fig.~\ref{figure_4}(d)-(f), a puller and neutral squirmer always swims in the positive $x$ direction for all separation distance while a pusher reverses its direction of motion at small gaps. Finally, the rotational velocities of the squirmer model are exhibited in Fig.~\ref{figure_4}(g)-(i); here the negative rotational velocity indicates that the swimmer tends to rotate counterclockwise (in Fig.~\ref{figure1}). A neutral squirmer rotates counterclockwise in all separation distances and its rotational speed monotonically increases. On the other hand, the direction of rotation for a puller and pusher can change depending on the separation distance: A puller (pusher) at large separation distance rotates clockwise (counterclockwise) however it rotates counterclockwise (clockwise) at smaller separation distances. At very thin gap size ($\delta\ll1$), the squirmer always swim slower next to a larger obstacle. The far field solutions for translational and rotational velocities given by Spagnolie \emph{et al.} \cite{Spagnolie_Lauga_2015} were plotted for pusher and puller squirmers and the predictions are accurate only when the swimmer is far from the wall. In particular, the far field solution could not predict the sign change in rotational velocities of the pusher and puller. This feature determines the trajectory of the swimmer as positive rotational velocity  compels the particle to stay close to the obstacle while the negative values force the particle to move away from the wall \cite{Wall_1}.
\subsection{Diffusiophoretic swimmer}
Considering a net repulsive hard-sphere excluded volume interaction between the product solute and the particle, a diffusiophoretic swimmer, far away from a boundary, translates in the direction of its orientation vector. Furthermore, since the driving force for propulsion is originated from the asymmetric solute distribution, it is crucial to perceive the influence of a curved boundary on both solute distribution and hydrodynamics of the problem. The net contribution of these two effects control the path of the swimmer.  At sufficiently small rate of reaction relative to the solute diffusion, the solute distribution around the particle is a function of the active coverage $\theta_{\rm{cap}}$, particle orientation angle $\Xi$, the separation distance to the obstacle $\delta$ and the size (curvature) of the wall $\kappa$.

\begin{figure}
\centering
\includegraphics[width=0.7 \textwidth]{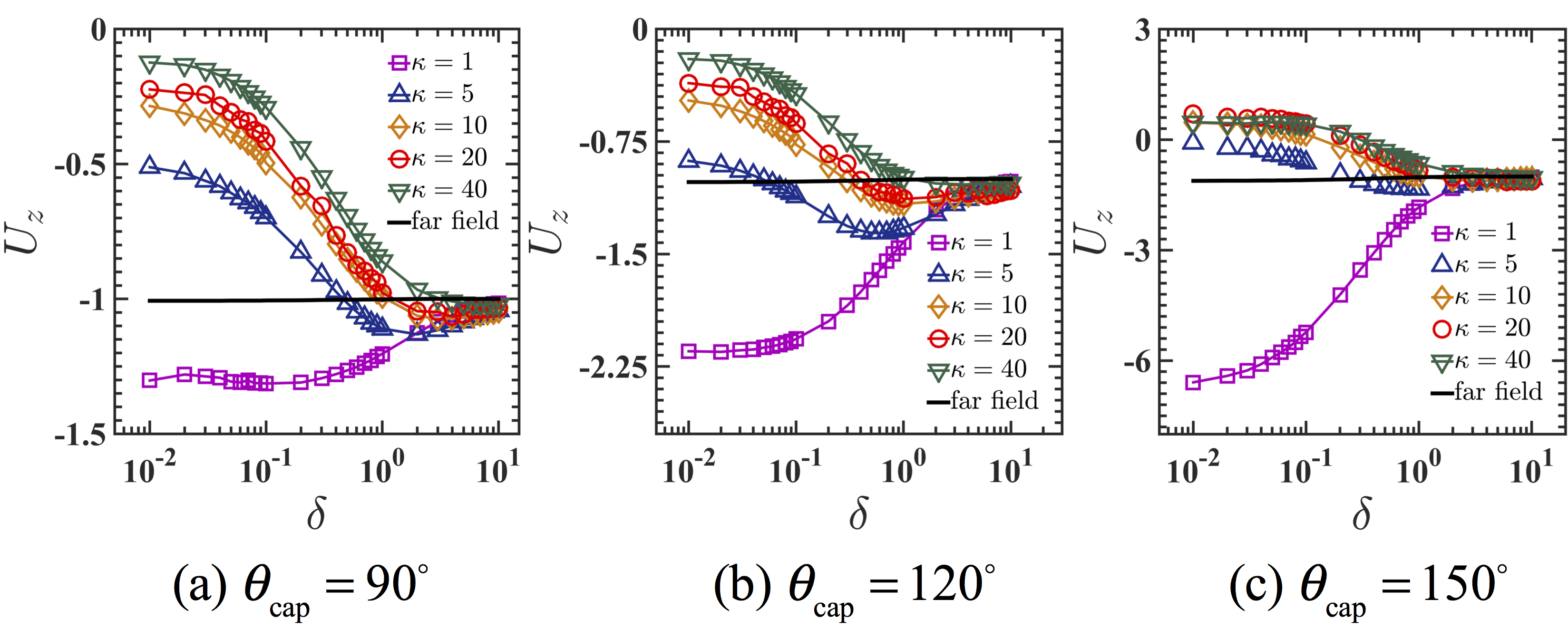}     	    
\caption{\footnotesize{The non-dimensional velocities of a diffusiophoretic swimmer as a function of non-dimensional edge-to-edge separation distance $\delta$ approaching a curved wall (with various degree curvature) with the tilt angle of $\Xi = 180^{\circ}$ for $(a)$ $\theta_{cap}=90^{\circ}$, $(b)$ $\theta_{cap}=120^{\circ}$ and $(c)$ $\theta_{cap}=150^{\circ}$}.} 
\label{figure_5}
\end{figure}

Regarding axisymmetric motion of a phoretic swimmer, i.e. $\Xi=180^{\circ}$ and $\Xi=0^{\circ}$, the swimmer does not rotate as the net hydrodynamic force is symmetric with respect to its center of mass. Moreover, it was previously shown \cite{Michelin2,Wall_1,Nima_3} that a diffusiophoretic swimmer approaches and moves away from an obstacle (or flat wall) with different velocities. Fig.~\ref{figure_5} exhibits the axisymmetric motion of a diffusiophoretic swimmer with various coverages next to a curved wall. In all cases, the solute concentration gradient around the swimmer increases as it moves toward the curved obstacle and thus it is initially expected that the particle swims faster. However, as the separation distance decreases the hydrodynamic drag increases drastically (due to lubrication forces) and therefore the balance between these two effects give rise to the observed behavior.  For a Janus swimmer (see Fig.~\ref{figure_5}(a)) with $\theta_{cap}=90^{\circ}$, the swimmer velocity drops at small separation for all obstacle sizes except for $\kappa=1$, where the size of the obstacle is equal to the swimmer. In this particular case ($\kappa=1$), the rate of increase of the hydrodynamic drag resistance is weaker relative to the rate of increase to the propulsive force and hence the swimmer translates faster even at small distance from the obstacle. For larger cap sizes of $\theta_{cap}=120^{\circ}$ (see Fig.~\ref{figure_5}(b)), a similar trend can be observed for the swimmer, however, the phoretic effect for larger cap is bolder. When the cap size becomes even larger, the swimmer tends to reverse its direction of motion (see Fig.~\ref{figure_5}(c)) for all obstacle sizes (except $\kappa=1$), suggesting that there is a separation distance in which the swimmer reaches a complete stationary state. Utilizing the method of reflections for concentration field, a far field solution can be found for axisymmetric motion of a phoretic swimmer against an inert obstacle (see Appendix B for detail). The far field solution is plotted in Fig.~\ref{figure_5} in black solid lines. The far field solution has a very weak dependency on the size of the obstacle and based on the far field solution, the phoretic velocity of the swimmer scales as $\Delta^{-5}$ when the swimmer is far from the obstacle.  It is noteworthy to emphasize that for the axisymmetric motion of a phoretic swimmer, the swimmer moves away from the wall much faster relative to the situation where it traverses toward the wall. As discussed previously, this counterintuitive behavior relies on the mechanism involved in the propulsion as the concentration fields vary considerably different in these two situations. In general, as a phoretic particle approaches a curved wall, the concentration gradient around the swimmer and the hydrodynamic drag increase and the balance between these two give rise to the final swimming velocity of the particle. However, the concentration field of the solute highly depends on the particle orientation and brings about the difference. Additionally, as the size of the wall increases, the lubrication force becomes much more significant and  therefore at small distances, the swimming velocity of the particle would be smaller for flatter obstacles. This figure evidently depicts one of the important messages of this study: the larger the obstacle, the greater possibility of trapping of the swimmer.
\begin{figure}
\centering
\includegraphics[width=0.7 \textwidth]{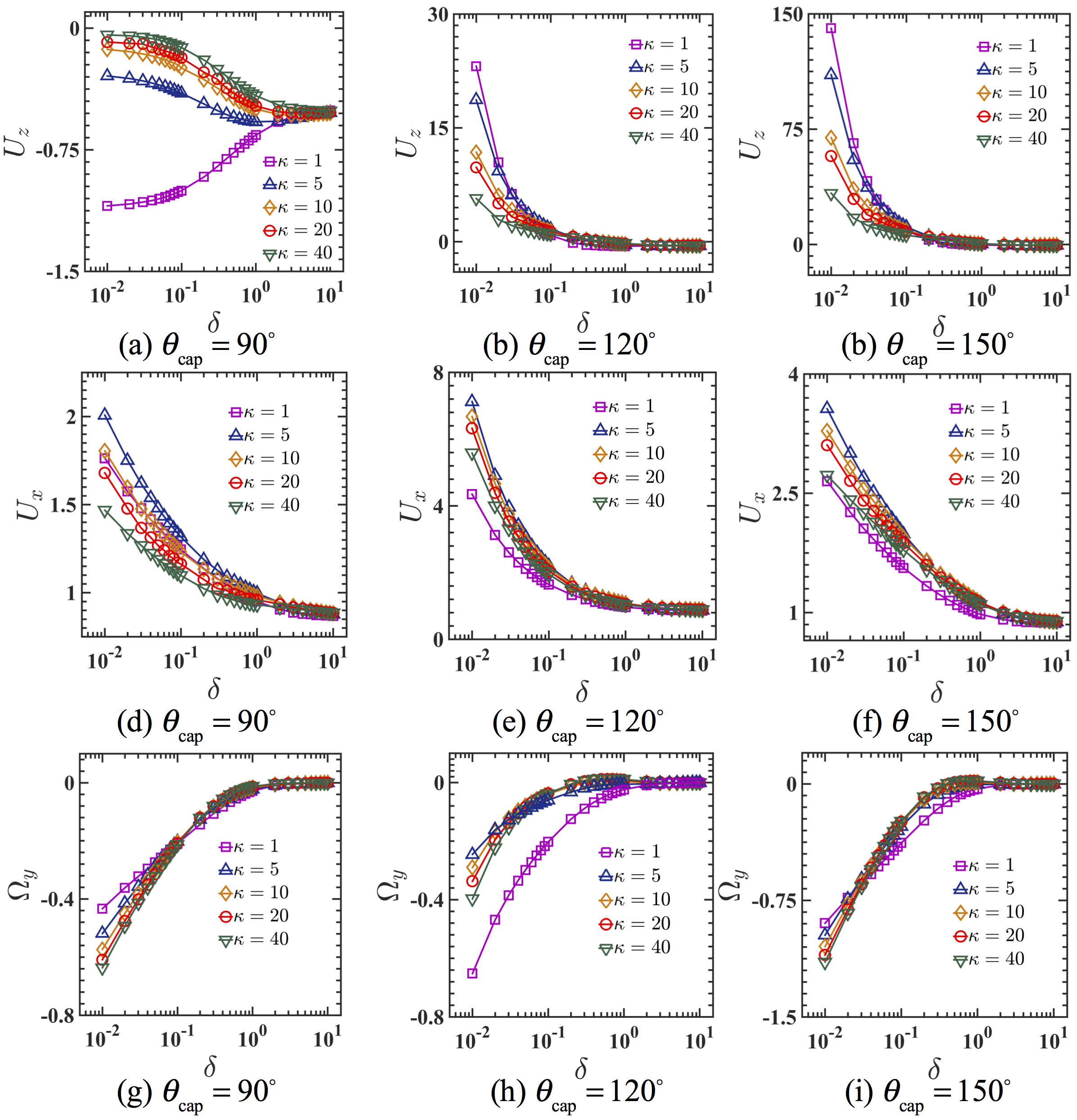}     	    
\caption{\footnotesize{The non-dimensional velocities of a diffusiophoretic swimmer as a function of non-dimensional edge-to-edge separation distance $\delta$ approaching a curved wall (with various degrees of curvatures) with a tilt angle of $\Xi = 120^{\circ}$ for $\theta_{cap}=90^{\circ}$, $\theta_{cap}=120^{\circ}$ and $\theta_{cap}=150^{\circ}$}.} 
\label{figure_6}
\end{figure}

For asymmetric motions, a phoretic swimmer rotates and translates simultaneously. Fig.~\ref{figure_6} illustrates phoretic velocities of a swimmer with coverages of $90^{\circ}$, $120^{\circ}$ and $150^{\circ}$ as a function of non-dimensional separation distance for arbitrary tilt of $\Xi=120^{\circ}$. As shown in Fig.~\ref{figure_6}(a)-(c), the swimming velocity in the $z$ direction (normal to the wall) is negative for $\theta_{cap}=90^{\circ}$ for all obstacle sizes and the speed decreases as the swimmer approaches the wall. By contrast, for a large cap size swimmer, e.g. $\theta_{cap}=120^{\circ}$ and $\theta_{cap}=150^{\circ}$, $U_z$ is negative only when the swimmer is very far from the wall and it switches sign to become positive as the swimmer moves closer. This particular trend indicates that a phoretic swimmer with large cap size and at certain tilt angles can gain an equilibrium distance in which its velocity in the $z$ direction is zero and thus can only swim in a direction parallel to the obstacle. Moreover, as the sign of velocity in $z$ direction becomes positive for smaller separation distances, the swimmer is repelled from the obstacle. In all cases however, the dynamics of phoretic swimmer are considerably slower next to a large obstacle as the lubrication force is much stronger in this case. For motions parallel to the obstacle (shown in Fig.~\ref{figure_6}(d)-(f)), the swimmer translates faster (in $x$ direction) at smaller separation distances as the swimming thrust (propulsive force) becomes stronger when swimmer is next to a no-slip wall. Regarding rotation around the third ($y$) axis, a phoretic swimmer with the coverage of $\theta_{cap}=90^{\circ}$ has a negative rotational velocity irrespective of the obstacle size and separation distance (see Fig.~\ref{figure_6}(g)) indicating that the swimmer tends to rotate counterclockwise to align its orientation vector with the normal vector of the obstacle. This evidently causes the swimmer to be repelled from the wall. On the other hand, for larger cap angles, the swimmer has a positive torque at large distances and it switches sign as the separation distance becomes smaller (see Fig.~\ref{figure_6}(h) and (i)). This trend also indicates that a phoretic swimmer with a specific cap angle at a given separation distance can reach a state where it does not rotate and only translates. Given this description, a phoretic swimmer close to an obstacle may reach a distance in which both $U_z$ and $\Omega_y$ become zero while $U_x>0$, and therefore, the swimmer would translate ``only'' parallel to the obstacle at an equilibrium distance. At very small separation distances however, the rotational velocity becomes negative and hence the swimmer rotates counterclockwise. The obstacle size significantly influences the rotational dynamics of the phoretic microswimmer: for small obstacle size ($\kappa=1$), the swimmer always rotates counterclockwise for all cap angles however as the obstacle size increases, the rotational velocity switches sign and becomes positive. Additionally, the results shown in Fig.~\ref{figure_6} (g)-(i) indicate that at very small distances, the magnitude of rotational dynamics is always faster next to a larger obstacle.
\subsection{Trajectory}
To systematically investigate the trajectory of a microswimmer next to a curved wall, we assume the swimmer is initially located at a separation distance of $\delta =2$, its initial orientation vector is in the $x-z$ plane and Brownian motion is negligible. Furthermore, the particle remains in the same plane indefinitely as the component of rotational velocity causing the particle to leave this plane is identically zero. Also, to avoid numerical difficulties, an edge-to-edge separation distance of $\delta =0.01$ is considered below which we assume the swimmer is in contact with the obstacle. This cut-off is necessary for two reasons: first, there are no extra repulsive force (e.g. steric repulsion) in our model as in principle the lubrication force must be sufficient to hinder particle impact and secondly, the slip velocity argument is not a legitimate assumption in the limit where the interaction layer becomes comparable to the gap size.

The trajectory of a microswimmer next to a curved obstacle can be one of the four following scenarios: (a) collision where the swimmer comes into contact with the wall, (b) escape in which the swimmer either rotates counterclockwise (negative rotational velocity) and moves away or its orientation vector is directed in a way that it is repelled from the wall, (c) orbiting in which the swimmer orbits around the curved obstacle indefinitely and (d) stationary state in which the swimmer reaches an equilibrium distance for which all components of translational and rotational velocities are identically zero and hence the swimmer remains at a standstill location indefinitely. Among these various trajectories, the most noticeable ones are the orbiting and stationary trajectories as they may have practical applications. Below these two types of trajectories for two classes of swimmers are discussed and the circumstances under which the microswimmer orbits or reaches a stationary state are fully revealed. 
\subsection{Squirmer}
\begin{figure}
\centering
\includegraphics[width=0.5 \textwidth]{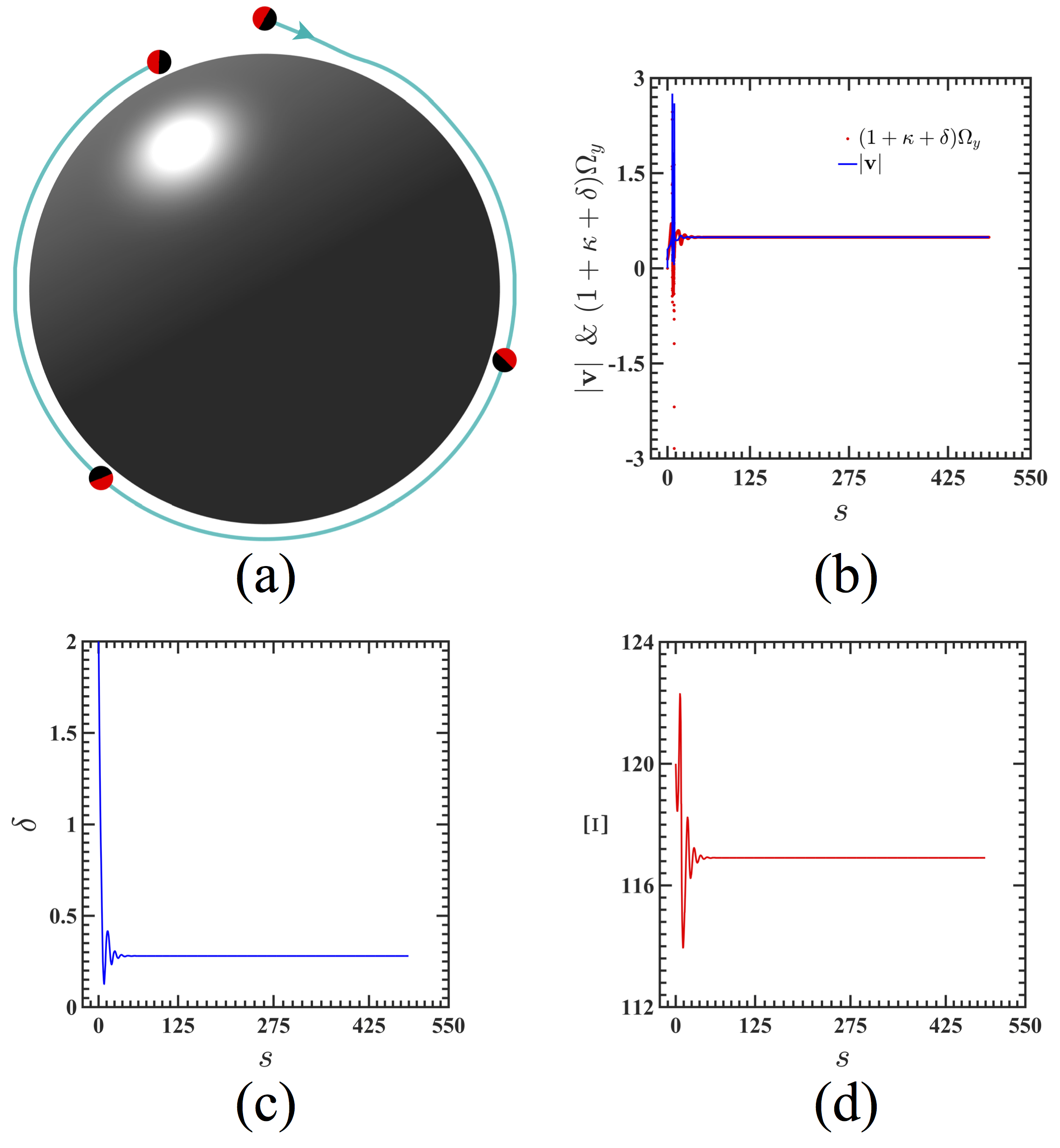}     	    
\caption{\footnotesize{The asymmetric motion of a puller ($\gamma =8$) against a curved wall ($\kappa=20$). (a) the trajectory, (b) the velocity magnitude $\left| {\bf{v}} \right| = \sqrt {v_x^2 + v_z^2}$, (c) gap distance $\delta$ and (d) local tilt $\Xi$ as a function of arclength $s = \int_0^t {\sqrt {v_x^2 + v_z^2} dt'}$. The equilibrium distance is $\delta_{eq}=0.28$ and the tit angle is $\Xi_{eq}=117^{\circ}$. }} 
\label{figure_7}
\end{figure}

Considering an initial tilt angle of $\Xi=180^{\circ}$ for a squirmer next to a curved wall, a neutral squirmer and a puller have negative velocity regardless of obstacle size and therefore they both ultimately come into contact with the obstacle. On the other hand, the magnitude of velocity  for a pusher decreases and below a separation distance it becomes positive (e.g. the swimmer switches its direction of motion). This particular behavior of the pusher permits it to reach an equilibrium distance in which it stays still. This equilibrium distance marginally depends on the size of the obstacle (greater equilibrium distance for larger obstacle size). By contrast, a puller translates faster as it approaches the obstacle and thus it ultimately hits the wall regardless of the obstacle size.
For an oblique incident, a puller at certain tilt angles and strength can orbit around the obstacle. This is due to the fact that a puller as shown in Fig.~\ref{figure_4}(h) can have a positive rotational velocity. The magnitude of the rotational velocity is boosted as the puller strength becomes greater. While a puller orbits a curved obstacle, it holds a constant distance relative to the obstacle, its relative orientation with respect to the curved obstacle is also fixed and thus the magnitude of (positive) rotational velocity of the swimmer must be sufficiently large such that the swimmer holds its distance fixed relative to the wall. Fig.~\ref{figure_7}(a) exhibits the orbiting trajectory of a puller around a curved obstacle of $\kappa=20$. Inasmuch as the trajectory of the puller is circular, the magnitude of translational velocity $\left| {\bf{v}} \right|$ must be equal to the rotational velocity $\Omega_y$ times the radius of orbit $1+\kappa+\delta$. The magnitude of translational velocity and rotational velocity times the radius of orbit are plotted against the arclength $s$ in Fig.~\ref{figure_7}(b). Evidently, the two values are not initially identical as the particle is not yet in its orbiting state however they eventually become identical later on when the swimmer is in its orbit. Moreover, as shown in Figs.~\ref{figure_7}(c) and (d), the puller reaches an equilibrium distance and tilt angles (in the co-moving coordinate) with respect to the obstacle. The strength of the puller is another parameter controlling the orbiting trajectory. The analysis indicates the equilibrium distance and the tilt angle depend on the strength of the puller and the critical strength beyond which the puller starts to orbit increases as the size of obstacle decreases (e.g. $\gamma_{\rm{critical}}=5$ for $\kappa=40$, $\gamma_{\rm{critical}}=7$ for $\kappa=20$ and $\gamma_{\rm{critical}}=14$ for $\kappa=10$). The orbiting trajectory would occur for even stronger puller $\gamma>\gamma_{\rm{critical}}$ however the values of the equilibrium distance (local tilt angle) would be smaller (greater).  On the other hand, a neutral squirmer always possesses a negative rotational velocity (Fig.~\ref{figure_4}(g)) against an obstacle and thus only two possible trajectories can occur for a neutral squirmer, either it rotates until its relative tilt becomes less than $90^{\circ}$ allowing it to escape or alternatively it collides with the wall before it reorients. Likewise, a pusher cannot orbit around an obstacle as its rotational velocity switches sign from negative to positive values as it approaches an obstacle (Fig.~\ref{figure_4}(i)). This trend leads again to only two types of trajectories for a pusher, for initial tilt angles of $\Xi \sim 90^{\circ}$ the swimmer reorients and escapes from the obstacle and for greater values of the initial tilt angles, the swimmer collides with the wall. 
\subsection{Diffusiophoretic swimmer}
To explore all possible scenarios for the phoretic swimmer against curved obstacles, we change particle surface coverage, particle orientation vector and the size of the obstacle. For axisymmetric motion where the initial angle of the swimmer was set to $\Xi = 180^{\circ}$ the swimmer moves towards the obstacle without any rotation. In this case, the particle can either hit the obstacle or reach a stationary state depending on its surface coverage. For small surface coverages, the swimmer always impacts the obstacle regardless of the obstacle size, nonetheless, the particle can reach a stationary state with an equilibrium distance for larger surface coverages. The critical surface coverage above which the particle is always stationary has a monotonic trend with the obstacle size, these critical cap angles are ${\theta _{{\rm{cap}}}} = {163^ \circ },~{151^ \circ },~{148^ \circ },~{146^ \circ }$ and ${145^ \circ }$ for the obstacle sizes of $\kappa=1,~5,~10,~20$ and $40$, respectively. Moreover, the equilibrium distance to the obstacle in which the swimmer with a critical cap size comes to a complete stop depends on the obstacle size as  ${\delta _{\rm{eq}}} = 0.023,~0.043,~0.041,~0.028$ and $0.024$ for the obstacle sizes of $\kappa=1,~5,~10,~20$ and $40$, respectively. This indicates that the stationary state can occur for a greater number of cap angles for a swimmer next to a larger obstacle relative to a smaller one. 

In asymmetric incidents, where the initial orientation angle of the particle is small i.e. $\Xi<90^{\circ}$, the particle is repelled from any obstacle  with arbitrary size as the swimming velocity in the $z$ direction is positive. For larger initial tilt $\Xi>90^{\circ}$, the direction of rotation is the key parameter in determining the trajectory of the swimmer. For small surface coverages, the rotational velocity of the particle is negative and therefore the particle rotates counterclockwise (negative rotational velocity) and hence there are only two possible trajectories: for small initial orientation $\Xi \sim 90^{\circ}$, the particle reorients and thus escapes, and for larger initial tilt angle $\Xi \sim 180^{\circ}$, it reaches the edge-to-edge separation distance of $\delta = 0.01$ before it has a chance to reorient and thus it hits the obstacle. For larger surface coverages, the rotational velocity of the particle can become positive (see Fig.~\ref{figure_6}(h) and (i)) and this permits the particle to skim along the obstacle to achieve a closed orbit. Notice possessing a positive rotational velocity is only a necessary condition for reaching a full circular trajectory. As a matter of fact, the magnitude of the particle rotational velocity must be large enough to keep up with the local curvature of the obstacle so that it remains in its equilibrium distance and relative orientation angle. In many scenarios, the particle has a positive rotational velocity but cannot keep up with the wall curvature. Thus it only skims the surface partially and  ultimately escapes. For very large surface coverage, the magnitude of positive rotational velocity is so large that it allows the particle to reach a local axisymmetric incident and thus it comes to a complete stationary state. Inasmuch as the magnitude of the rotational velocity is greater for larger obstacles, there is a greater chance for both orbiting and stationary trajectories. 
\begin{figure}
\centering
\includegraphics[width=0.6 \textwidth]{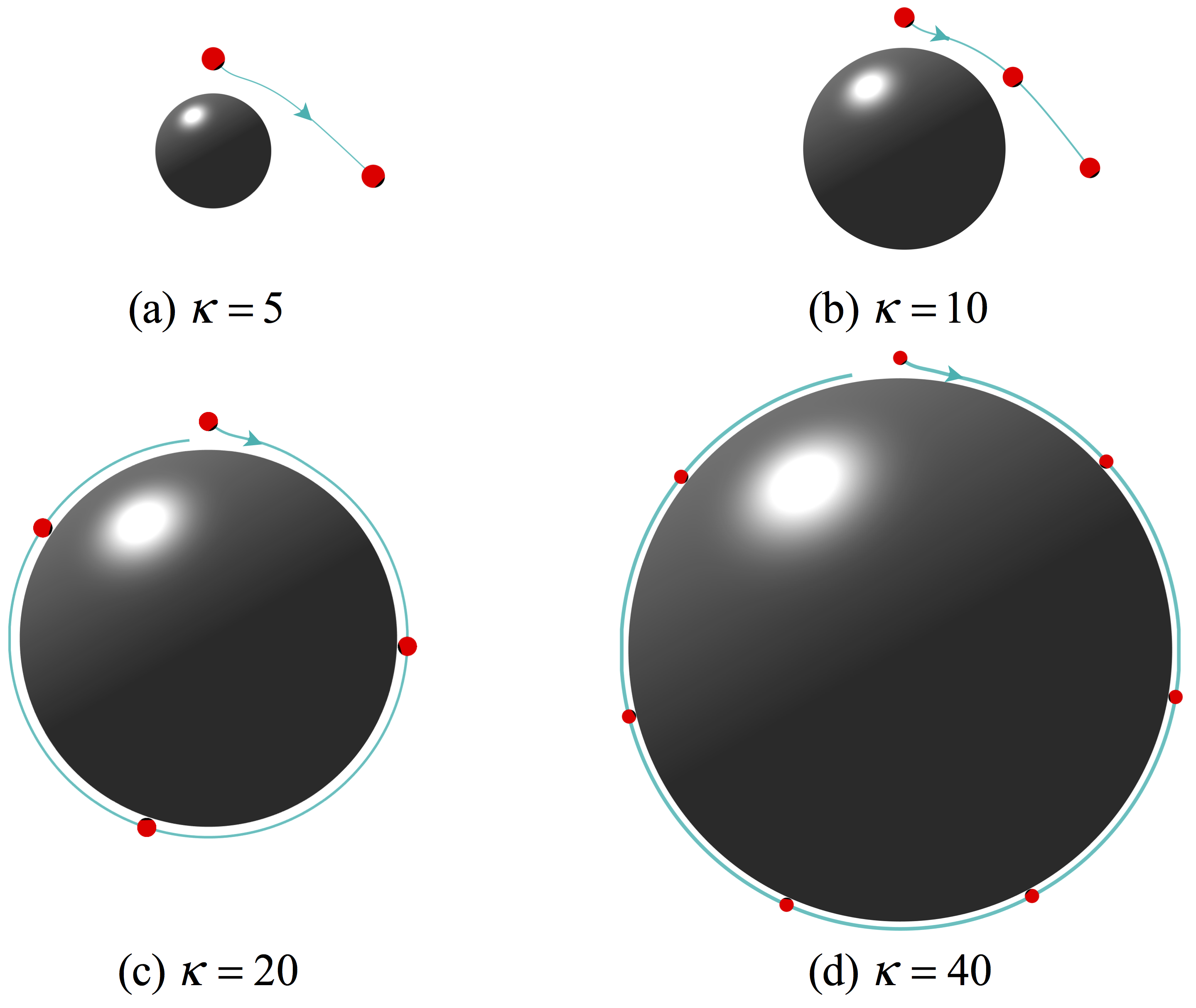}     	    
\caption{\footnotesize{The trajectories of a phoretic swimmer with initial orientation angle of $\Xi=140^{\circ}$ and $\theta_{\rm{cap}}=140^{\circ}$ near various obstacle sizes.}} 
\label{figure_8}
\end{figure}
Fig.~\ref{figure_8} demonstrates the trajectories of a phoretic swimmer with initial orientation angle of $\Xi=140^{\circ}$ and $\theta_{\rm{cap}}=140^{\circ}$ near various obstacle sizes. While the swimmer orbits for obstacle sizes of $\kappa=20$ and $40$, it fails to completely orbit for smaller obstacle size of $\kappa= 5$ and $10$. This particular trend is due to the fact that for small obstacles, the magnitude of rotational velocity is not large enough to let the particle possess a fixed relative orientation angle with respect to the obstacle and thus it exposes its active surface to the wall and thereafter escape from it. From geometrical point of view, a swimmer can rotate clockwise in a circular path if its translational speed becomes equal to its rotational velocity magnitude times the radius of orbit, i.e. $\left| {\bf{v}} \right| = \left( {1 + \kappa  + \delta } \right)\left| {\bm{\Omega}}  \right|$. For large obstacles, this criteria can be satisfied as the magnitude of the rotational velocity is large enough to allow orbiting however for smaller obstacle, the magnitude of rotational velocity never reaches the critical value where the previous relation is satisfied at any separation distance and particle active coverage and hence no orbiting trajectory can ever be observed for smaller obstacles. Given these considerations, we performed numerous simulations (various active coverage and initial orientations) to sketch the complete picture. The phase diagrams of a phoretic swimmer against obstacles of $\kappa=20$ and $\kappa=40$ with initial gap distance of $\delta=3$ are shown in Fig.~\ref{figure_9}(a) and (b) respectively. As evident in both cases the swimmer can orbit for large active coverage and the orbiting and stationary regimes extend with obstacle size. By contrast, orbiting was not observed for smaller obstacle sizes ($\kappa \le10$) as in this case the magnitude of positive rotational velocity is not large enough to allow the particle to stay in the orbit.  
\begin{figure*}
\centering
\includegraphics[width=0.9 \textwidth]{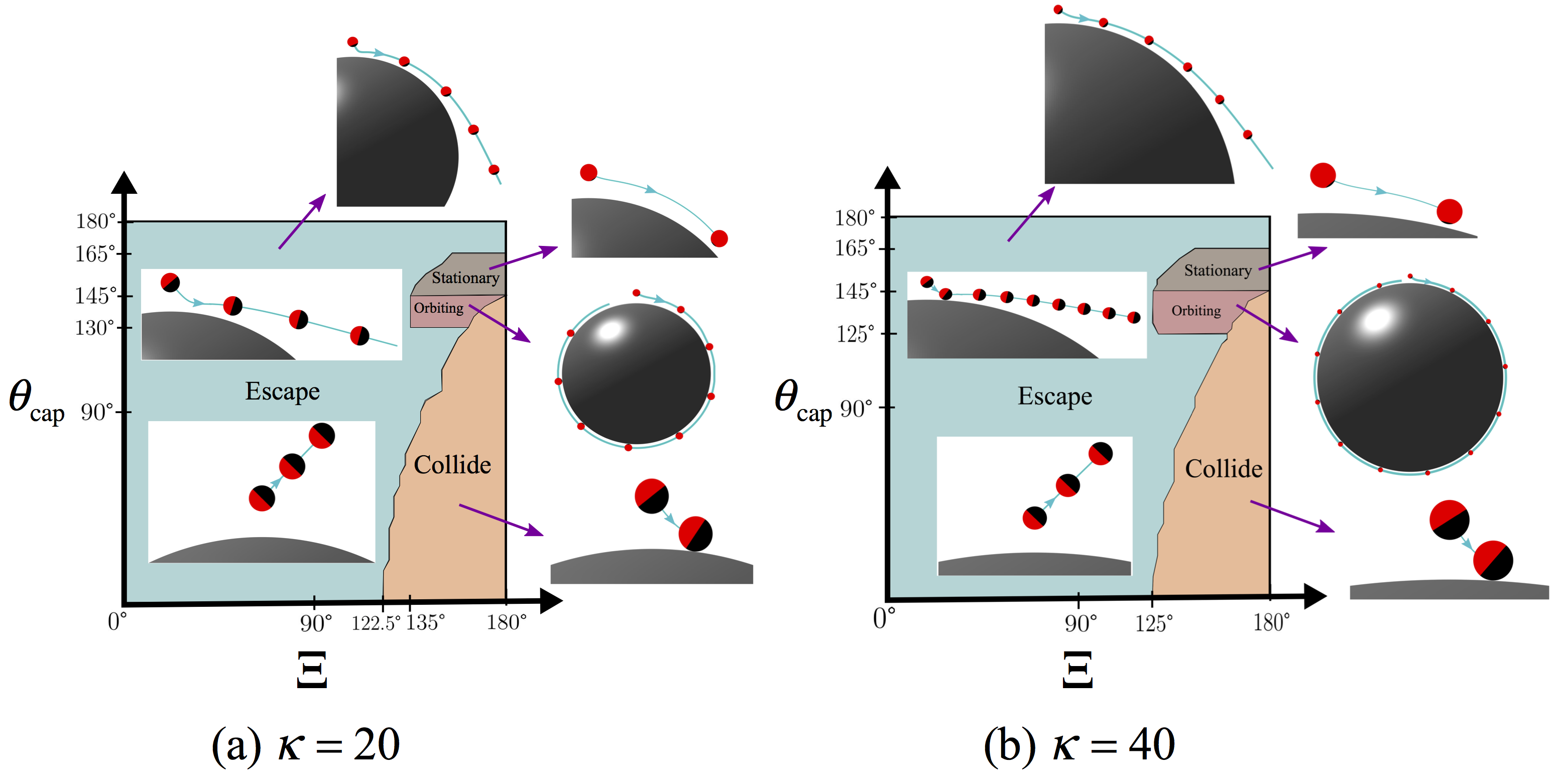}     	    
\caption{\footnotesize{Phase diagrams of a phoretic swimmer next to a curved wall for (a) $\kappa=20$ and (b) $\kappa=40$. See videos in Supplementary Information for various trajectories.}} 
\label{figure_9}
\end{figure*}
\section{Summary}
A hydrodynamic model based on Reynolds reciprocal theorem (RRT) was developed to study the dynamics and trajectories of a microswimmer next to a spherical obstacle. The model can be equally applied to both biological and artificial microswimmers as long as the propulsion mechanism can be expressed as a slip velocity on the swimmer surface. Moreover the proposed model is uniformly valid in both far and near fields as it accounts for the exact hydrodynamic interaction of the swimmer near curved spherical obstacles. Generally, microswimmers are force and torque free and regardless of the swimmer types and the mechanism for the propulsion, the flow field around an isolated microswimmer (far from boundaries) is a force dipole up to the leading order. The dipolar field was used in conjunction with the method of images to describe the attraction and trajectories of biological swimmers to both planar and curved boundaries. This approach is accurate when the microswimmer is far from the obstacle, nonetheless it immediately loses its accuracy when the microswimmer is close to the obstacle where higher order modes become as important as the force dipole mode. Thus the behavior of different classes of microswimmers would be substantially distinct close to the boundaries as their detailed flow fields are not similar. On the other hand, since a number of interesting trajectories of microswimmers close to curved boundaries, including orbiting and stationary, take place in a very close proximity to the boundary we took advantage of a uniformly valid hydrodynamic approach which can self-consistently take into account the contribution of all modes. Although the image solution can be employed similarly to account for the contribution of higher order modes next to a curved obstacle, the algebra quickly becomes cumbersome for higher order modes. By contrast, RRT allows us to come up with a compact expression for swimming translational/rotational velocities of a microswimmer next to a curved wall as a function of the classically known solutions of translation and rotation of two spheres with arbitrary radii in the Stokes flow regime. 

Two conventional models, so-called squirmer and phoretic, utilized for describing the dynamics of biological and man-made microswimmers respectively, were considered. For the squirmer model, our analysis indicates that a puller can indeed orbit curved obstacles and for each obstacle size, there is a critical strength for a puller in which the swimmer can orbit and this critical strength is greater for smaller obstacle. By contrast a neutral squirmer and a pusher cannot orbit and they either escape or hit the wall. Regarding a phoretic swimmer against a curved obstacle, the analysis suggests that the particle can orbit only when its active coverage is large and as the coverage becomes smaller the particle either escapes or hits the obstacle. Additionally, the orbiting trajectory can take place for more active coverages for larger obstacle and as the obstacle size decreases, the particle can orbit for a smaller range of active coverages and for small obstacle size the orbiting can never occur. 

The proposed solution in this study can be utilized to understand the dynamics and trajectories of biological microswimmers. The accumulation of bacteria next to the wall brings about biofilm formation and understanding the dynamics of the bacteria under the confinement can permit designing systems where biofilm formation can be hindered or accelerated. Additionally, inasmuch as the detailed flow field generated by microbes are not similar, curved stationary obstacles can be used for sorting them. On the other hand, a similar strategy can be employed for guidance of synthetic microswimmers. 
  
The machinery introduced in this paper can be applied to a number of other problems, e.g. the motion of microswimmers near stress free boundaries at small capillary numbers (negligible interface deformation) to probe the motion of microswimmers next to a spherical bubble (the auxiliary problems in RRT must be replaced in this case \cite{Subramanian}). Moreover, the model can be readily extended to account for the inclusion of a non-uniform body force \cite{teubner} as some bacteria and Janus particles might be heavy and thus the effect of gravity might be important. The model can also be used for cases where there is a slip (chemiosmotic) flow on the wall. This might be relevant for the motion of Janus phoretic swimmers as the slip flow generated at the wall can modify the trajectory of the swimmer  \cite{Simmchen_2016,Uspal_2016}. The extension of the above model to consider Brownian motion of the swimmer is possible as recently shown by Mozaffari \emph{et. al} \cite{Wall_2}. The Brownian motion of a swimmer can randomize the direction of motion of the swimmer and thus gives rise to the escape of the swimmer from closed orbit. As shown for the case of a planar wall, at high P\'eclet numbers (defined as $Pe\sim \mu a_1^2{{{U_c}} \mathord{\left/{\vphantom {{{U_c}} {{k_B}T}}} \right. \kern-\nulldelimiterspace} {{k_B}T}}$), the swimmer always traces out the deterministic limit and as $Pe$ decreases, the probability of escape increases. In principle, one can still apply Eq.~\ref{compact} for more complex situations where the shape of the swimmer is not spherical or the obstacle radius of curvature is not uniform \cite{Wykes_2017}. Nonetheless, the procedure cannot be followed analytically as the detailed flow fields of the test problem is not known analytically and thus a purely numerical approach (e.g. boundary element method) is required to systematically investigate the influence of geometry of the swimmer and the curved obstacle. 

\section*{Acknowledgment}
U.M.C.-F. was supported by National Science Foundation grants CBET -- 1055284 and EPS -- 1010674. P.G.D.-H was supported by Puerto Rico NASA Space Grant Fellowship --NNX15AI11H.\\

\section*{Appendix A}\label{appendix}
 In this section, we recapitulate the velocity field around a single squirmer and a catalytically active spherical swimmer far away from a boundary. The hydrodynamics at low Reynolds number for a single spherical particle can be written as,
\begin{align}
&\nabla  \cdot {\bf{v}} = 0,\\
&{\nabla ^2}{\bf{v}} - \nabla p = 0,
\end{align}
with the boundary condition defined as,
\begin{align}
&{\left. {{\bf{v}} = {{\bf{v}}_s} + {\bf{U}}} \right|_{r = 1}},\label{bc1}\\
&\mathop {\lim }\limits_{r \to \infty } {\bf{v}} \to 0,\label{bc2}\\
&{\bf{F}}_T= \oint_{{S_c}} { {{\bf{e}}_n \cdot {\bm{\sigma}}~ dS}}  = 0. \label{bc3}
\end{align}
Notice $r$ is the non-dimensional radial distance in the spherical coordinate located at the particle center and inasmuch as the motion is axisymmetric, there is no net torque and consequently no rotation for the swimmer.  To solve the hydrodynamics, we utilize the stream function approach in spherical coordinates ($r,\theta,\phi$), which is a conventional tool to solve axisymmetric flow problems in Stokes flow regime,
\begin{align}
&{v_r} =  - \frac{1}{{{r^2}\sin \theta }}\frac{{\partial \Psi }}{{\partial \theta }},\\
&{v_\theta } =   \frac{1}{{r\sin \theta }}\frac{{\partial \Psi }}{{\partial r}},
\end{align}
where the stream function can be governed directly by solving:
\begin{eqnarray}
{{\bf{E}}^4}\Psi  = 0,\label{sf}
\end{eqnarray}
where ${{\bf{E}}^4}$ is an operator defined according to:
\begin{align}
&{{\bf{E}}^4} = {{\bf{E}}^2}{{\bf{E}}^2},\\
&{{\bf{E}}^2} = \left[\frac{{{\partial ^2}}}{{\partial {r^2}}} + \frac{{\sin \theta }}{{{r^2}}}\frac{\partial }{{\partial \theta }}\left(\frac{1}{{\sin \theta }}\frac{\partial }{{\partial \theta }}\right)\right].
\end{align}
The general solution of Eq.~\ref{sf} will be:
\begin{eqnarray}
\Psi  = \sum\limits_{n = 1}^\infty  {\left[{a_n}{r^{n + 3}} + {b_n}{r^{ - n + 2}} + {c_n}{r^{n+1}} + {d_n}{r^{ - n}}\right]{C_{n+1}^{-\frac{1}{2}}}(\eta)} ,
\end{eqnarray}
where $a_n$, $b_n$, $c_n$ and $d_n$ are constants required to be evaluated by the boundary conditions, $\eta=\cos\theta$ and $C_n^{-\frac{1}{2}}(\eta )$ are the Gegenbauer polynomials of order $n$ and degree $-\frac{1}{2}$. By applying the far field boundary condition (Eq.~\ref{bc2}), one can instantly realize that,
\begin{align}
&a_n=0,\\
&c_n=0.
\end{align}
The boundary condition on the swimmer surface relies on the mechanism that is utilized for the propulsion nonetheless the slip velocity at the swimmer surface can be cast into the following form
\begin{align}
{\bf{v}}_s = {{\bf{e}}_\theta }\sum\limits_{n = 1}^\infty  {A_n \frac{n(n+1)}{{{{(1 - {\eta ^2})}^{\frac{1}{2}}}}}} C_{n + 1}^{ - \frac{1}{2}}(\eta),\label{sv}
\end{align}
where $A_n$ are the swimming modes that for the squirmer model are solely constant and $A_n = 0 $ for $n>2$ (see Eq.~\ref{squirmer-slip}). For the diffusiophoretic swimmer on the other hand, $A_n$ depend on the colloid coverage (see below for detail).  
Notice the swimmer translates along the $z$ axis with a constant velocity $U$ and thus 
\begin{eqnarray}
{\bf{U}} = U{{\bf{e}}_z} = U(\cos \theta {{\bf{e}}_r} - \sin \theta {{\bf{e}}_\theta }),
\end{eqnarray}
the stream function can be rewritten in terms of $\eta$ as,
\begin{align}
&{v_r} = \frac{1}{{{r^2}}}\frac{{\partial \Psi }}{{\partial \eta }} =  - \sum\limits_{n = 1}^\infty  {\left[{b_n}{r^{ - n}} + {d_n}{r^{ - n - 2}}\right]{P_n}(\eta )} ,\\
&{v_\theta } =  \frac{1}{{r{{(1 - {\eta ^2})}^{\frac{1}{2}}}}}\frac{{\partial \Psi }}{{\partial r}} =\nonumber \\
&  - \sum\limits_{n = 1}^\infty  {\left[(n - 2){b_n}{r^{ - n}} + n{d_n}{r^{ - n - 2}}\right]\frac{{C_{n + 1}^{ - \frac{1}{2}}(\eta )}}{{{{(1 - {\eta ^2})}^{\frac{1}{2}}}}}},
\end{align}
where we have used the following identity:
\begin{eqnarray}
\frac{{dC_{n + 1}^{ - \frac{1}{2}}(\eta )}}{{d\eta }} =  - {P_n}(\eta ),~~(n \ge 1),
\end{eqnarray}
where $P_n$ are the Legendre polynomials and since the slip velocity on the swimmer does not have $r$ component, the only contribution from the right hand side of the boundary condition given in Eq.~\ref{bc1} is from the swimming velocity and hence by using the orthogonality of the Legendre polynomials we have:
\begin{eqnarray}
\left\{ \begin{array}{l}
{b_1} + {d_1} =  - U,\\
{b_n} + {d_n} = 0,~~(n \ge 2),
\end{array} \right.\label{re1}
\end{eqnarray}
and similarly by applying the boundary condition on the swimmer surface along the $\theta$ direction and using the orthogonality of the Gegenbauer polynomials, we have,
\begin{eqnarray}
\left\{ \begin{array}{l}
{b_1} - {d_1} =  - 2U - 2{A_1},\\
(n - 2){b_n} + n{d_n} =  - n(n + 1){A_n},~~(n \ge 2),
\end{array} \right.\label{re2}
\end{eqnarray}
By solving Eq.~\ref{re1} and \ref{re2} simultaneously, one can deduce,
\begin{eqnarray}
\left\{ \begin{array}{l}
{b_1} =  - \frac{3}{2}U - {A_1},\\
{b_n} = \frac{{n(n + 1){A_n}}}{2},~~(n \ge 2),
\end{array} \right.
\end{eqnarray}
\begin{eqnarray}
\left\{ \begin{array}{l}
{d_1} = \frac{1}{2}U + {A_1},\\
{d_n} =  - \frac{{n(n + 1){A_n}}}{2},~~(n \ge 2),
\end{array} \right.
\end{eqnarray}
and finally the force free boundary condition, Eq.~\ref{bc3}, can be calculated as,
\begin{align}
\oint_{{S_c}} { {{\bf{e}}_n \cdot {\bm{\sigma}}~ dS} }= 2\pi \int_0^\pi  { - \frac{{\sin \theta }}{2}\frac{\partial }{{\partial r}}({{\bf{E}}^2}\Psi )d\theta  = \pi {b_1} = 0},
\end{align}
\begin{figure*}
\centering
\includegraphics[width=0.85 \textwidth]{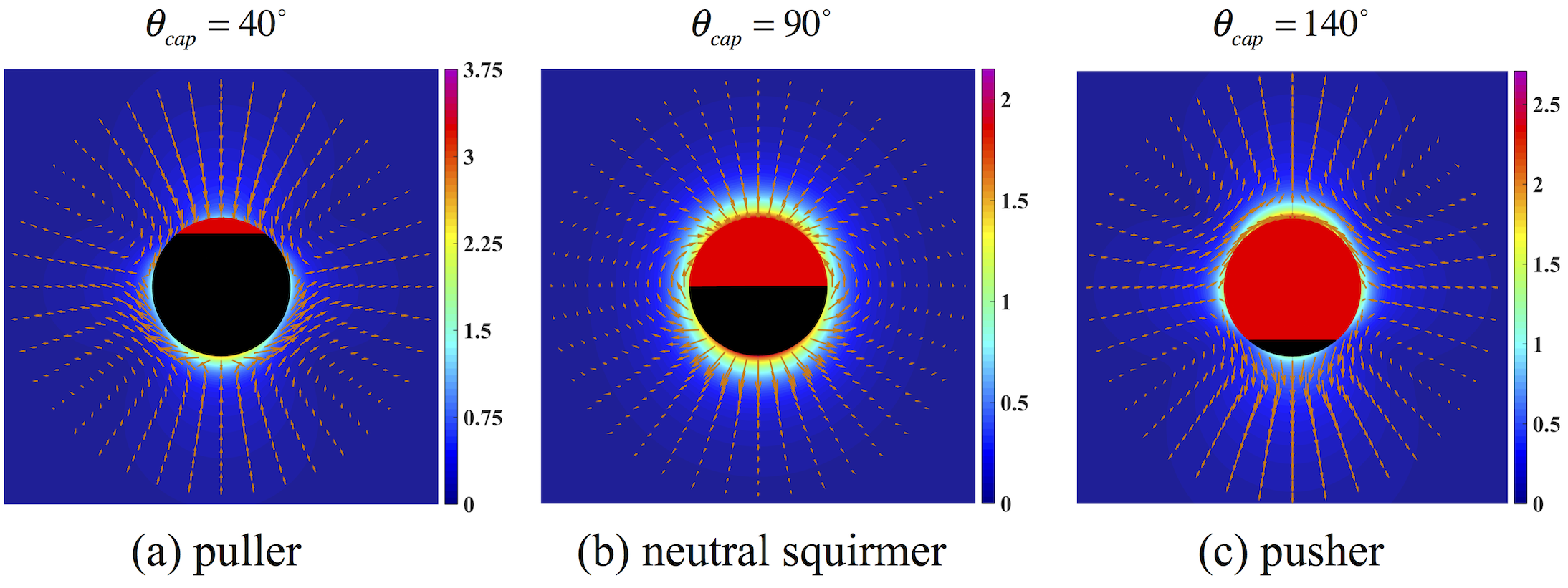}     	    
\caption{\footnotesize{The velocity field and non-dimensional Frobenius norm of total hydrodynamic stress tensor around a single phoretic swimmer far away from a boundary for $(a)$ $\theta_{cap}=40^{\circ}$ (puller), $(b)$ $\theta_{cap}=90^{\circ}$  (neutral squirmer) and $(c)$ $\theta_{cap}=140^{\circ}$ (pusher). Notice the swimmers towards the bottom (opposite to the reactive cap shown in red) in all three cases.}} 
\label{figure_1_appendix}
\end{figure*}
and therefore we conclude that:
\begin{align}
U=-\frac{2}{3}A_1,
\end{align}
\begin{eqnarray}
\left\{ \begin{array}{l}
{b_1} =  0,\\
{b_n} = \frac{{n(n + 1){A_n}}}{2},~~(n \ge 2),
\end{array} \right.
\end{eqnarray}
\begin{eqnarray}
\left\{ \begin{array}{l}
{d_1} = \frac{2}{3}A_1,\\
{d_n} =  - \frac{{n(n + 1){A_n}}}{2},~~(n \ge 2),
\end{array} \right. .
\end{eqnarray}
For squirmer model $A_1=B_1$ and $A_2={{{B_2}} \mathord{\left/{\vphantom {{{B_2}} 3}} \right.\kern-\nulldelimiterspace} 3}$ are constants and $A_n=0$ for $n>2$ and hence the velocity field is:
\begin{align}
&{v_r} =  - \frac{{2{B_1}}}{{3{r^3}}}\cos \theta  - \frac{{{B_2}}}{{2{r^2}}}\left( {1 - \frac{1}{{{r^2}}}} \right)\left( {3{{\cos }^2}\theta  - 1} \right),\label{vr_sq}\\
&{v_\theta } =  - \frac{{{B_1}}}{{3{r^3}}}\sin \theta  + \frac{{{B_2}}}{{2{r^4}}}\cos \theta \sin \theta \label{vt_sq},
\end{align}
there is no external force acting on the swimmer and hence no stokeslet (as $b_1$=0) and $\sim r^{-1}$ dependency in the the flow field. The next least decaying singularity that might be present by the activity of the swimmer is the force dipole attenuated as $\sim r^{-2}$ and from Eqs.~\ref{vr_sq} and \ref{vt_sq}, the far field flow of the pusher and puller can be well-described by a force dipole. For a neutral squirmer, $B_2 =0 $ and hence there is no force dipole and the flow field is that of source dipole (i.e. potential flow) reminiscent of electrophoretic motion of a charged particle under an external electric field \cite{Morrison_1970}. 

 For the phoretic swimmer, the solute distribution around a constant flux spherical swimmer with a symmetric cap size of $\theta_{cap}$ with respect to the $z$ axis can be governed by solving a Laplace equation in the spherical coordinates according to,
\begin{align}
& \nabla^2 n =0,\\
&n(r,\theta ) = \sum\limits_{n = 0}^\infty  {\frac{{{A_n}}}{{{r^{n + 1}}}}} {P_n}(\cos \theta ),\label{c1}\\
&{A_n} = \frac{{2n + 1}}{{2(n + 1)}}\int_0^{{\theta _{cap}}} {{P_n}(\cos \theta )\sin \theta d\theta } ,\label{c2}
\end{align}
with the first few terms calculated as,
\begin{align}
&{A_0} = \frac{1}{2}(1 - \cos {\theta _{cap}}),\\
&{A_1} = \frac{3}{8}(1 - \cos^2 {\theta _{cap}}),\label{A_1}\\
&{A_2} = \frac{5}{{12}}\cos {\theta _{cap}}(1 - \cos^2 {\theta _{cap}}).\label{A_2}
\end{align}
The slip velocity on the particle can be found according to
\begin{align}
&{\bf v}_s={\left. { - {\nabla _s}n} \right|_{r = 1}} = {{\bf{e}}_\theta }\sum\limits_{n = 0}^\infty  {{A_n}{{(1 - {\eta ^2})}^{\frac{1}{2}}}\frac{{\partial {P_n}(\eta )}}{{\partial \eta }}} \nonumber\\ 
 &= {{\bf{e}}_\theta }\sum\limits_{n = 0}^\infty  {{A_n}\frac{{n(n + 1)}}{{{{(1 - {\eta ^2})}^{\frac{1}{2}}}}}} C_{n + 1}^{ - \frac{1}{2}},\label{sv}
\end{align}
where in the above, we utilized the following identities to turn the phoretic slip velocity into the form introduced in Eq.~\ref{sv},
\begin{align}
&\frac{{\partial {P_n}(\eta )}}{{\partial \theta }} =  - {(1 - {\eta ^2})^{\frac{1}{2}}}\frac{{\partial {P_n}(\eta )}}{{\partial \eta }},\\
&\frac{{\partial {P_n}(\eta )}}{{\partial \eta }} =  \frac{{n(n + 1)}}{{(1 - {\eta ^2})}}C_{n + 1}^{ - \frac{1}{2}}(\eta ).
\end{align}

If the slip velocity relation given in Eq.~\ref{sv} truncated for $n \le 2$, then the velocity field around a single diffusiophoretic swimmer far from a boundary can be interpreted according to the squirmer model (Eq.~\ref{squirmer-slip}). In this case,  $A_1$ is always positive however $A_2$ (see Eqs.~\ref{A_1} and \ref{A_2}) is positive for $\theta_{cap}<90^{\circ}$, negative for $\theta_{cap}>90^{\circ}$ and zero for $\theta_{cap}=90^{\circ}$ and hence the flow field around the diffusiophretic swimmer can be puller, pusher and neutral squirmer respectively as shown in Fig.~\ref{figure_1_appendix}; a distinctive feature of small (puller) and large (pusher) cap sizes is the presence of the vortices induced by the swimmer stresslet formed in the front or back of the particle (see Fig.~\ref{figure_1_appendix} (a) and (c)). The Janus swimmer (neutral squirmer) on the other hand, does not have any vortices as it is determined based on the singularity solution of potential flow (see Fig.~\ref{figure_1_appendix} (b)). As shown in the above, the swimming velocity of a single swimmer far from a boundary or another swimmer is solely depends on the $A_1$ and it is always equal to ${\bf{U}}={2 \mathord{\left/{\vphantom {2 3}} \right.\kern-\nulldelimiterspace} 3}{A_1}{{\bf{n}}_p}$ regardless of the swimmer type (puller, pusher or phoretic swimmer) however the velocity, short and long time trajectories near an obstacle are substantially different for various types of swimmers as the reflective modes induced by the presence of the obstacle are completely different for various swimmers \cite{spagnolie_lauga_2012}. This fact is also evident for dilute and dense suspensions of the swimmer as their collective dynamics and the rheological properties highly depend on the flow field generated by their active elements  \cite{Saintillan_2008}.\\
The non-dimensional pressure field around the swimmer can be found by substituting the velocity components in the momentum equation (either $r$ or $\theta$ direction) and integrating to reach
\begin{eqnarray}
p = {p_\infty } - \sum\limits_{n = 1}^\infty  {\left[ {\frac{{2{b_n}}}{{{r^{n + 1}}}} - \frac{{2{n_{}}{d_n}}}{{(n + 3){r^{n + 3}}}}} \right]{P_n}(\eta )},
 \end{eqnarray}
 where $p_{\infty}$ denotes a reference pressure at a location far away from the particle center of mass. The non-zero components of the stress tensor can be also calculated according to
 \begin{align}
&{\sigma _{rr}} =  - p + 2\frac{{\partial {v_r}}}{{\partial r}},\\
&{\sigma _{\theta \theta }} =  - p + 2\left( {\frac{1}{r}\frac{{\partial {v_\theta }}}{{\partial \theta }} + \frac{{{v_r}}}{r}} \right),\\
&{\sigma _{r\theta }} = {\sigma _{\theta r}} = \left( {\frac{1}{r}\frac{{\partial {v_r}}}{{\partial \theta }} + \frac{{\partial {v_\theta }}}{{\partial r}} - \frac{{{v_\theta }}}{r}} \right),
\end{align}
 The Frobenius norm of the stress tensor, $\sqrt {\sum\limits_{i = 1}^2 {\sum\limits_{j = 1}^2 {\sigma _{ij}^2} } } $ where $1$ and $2$ denote $r$ and $\theta$ respectively, for a phoretic swimmer for different cap angles were shown in Fig.~\ref{figure_1_appendix}(a)-(c). While the region of high stress for the swimmer (a puller) with small coverage is close to its head, for high coverage swimmer (a pusher), the region of high stress is constrained to the tail of the swimmer (see Fig.~\ref{figure_1_appendix}(c)). The stress field is almost uniform with respect to the spherical polar angle $\theta$ for the case of $\theta_{cap}=90^{\circ}$ (neutral squirmer). 
 \section*{Appendix B}\label{Appendix B}
 Here, the detailed far field solution of a phoretic swimmer next to a spherical obstacle based on method of reflections is provided. Suppose there is a phoretic swimmer denoted by particle $1$ oriented axisymmetrically on top of a spherical obstacle denoted by $2$. The non-dimensional concentration field generated by the phoretic swimmer in the absence of the obstacle can be captured by Eqs.~\ref{c1} and \ref{c2} with respect to the local coordinate located at its center. This concentration field creates a dipolar field that must satisfy the zero flux condition on the obstacle surface and thus the concentration field around the obstacle due to the dipolar disturbance created by the swimmer would be
 \begin{eqnarray}
 {n_2} = {\zeta _0} + {\zeta _1}{r_2}\cos {\theta _2}\left( {1 + \frac{{{\kappa ^3}}}{{2r_2^3}}} \right),
 \end{eqnarray}
 where $r_2$ and $\theta_2$ denote the radial and polar angle components of the spherical coordinate located at the center of obstacle and ${\zeta _0}$ and ${\zeta _1}$ are defined as
\begin{eqnarray}
{\zeta _0} = \sum\limits_{n = 0}^\infty  {\frac{{{A_n}}}{{{\Delta ^{n + 1}}}}{P_n}( - 1),} \\
{\zeta _1} =  \sum\limits_{n = 0}^\infty  {\frac{{(n + 1){A_n}}}{{{\Delta ^{n + 2}}}}{P_n}( - 1)}. \label{zeta_1}
\end{eqnarray}
Likewise, the disturbance created by the presence of the obstacle is reflected back to the swimmer and thus it can be shown that the concentration field around the phoretic swimmer is 
\begin{eqnarray}
{n_1} = \sum\limits_{n = 0}^\infty  {\frac{{{C_n}}}{{r_1^{n + 1}}}{P_n}(\cos {\theta _1})}  + \frac{{{\zeta _1}{\kappa ^3}}}{{2{\Delta ^3}}}\left( {\Delta  - 2{r_1}\cos {\theta _1}} \right),
\end{eqnarray}
where
\begin{align}
{C_1} &= \frac{3}{8}(1 - \cos^2 {\theta _{cap}}) - \frac{{{\zeta _1}{\kappa ^3}}}{{{2\Delta ^3}}},\\
{C_n} &= \frac{{2n + 1}}{{2(n + 1)}}\int_0^{{\theta _{cap}}} {{P_n}(\cos \theta )\sin \theta d\theta },~~(n\ne1).
\end{align}
Considering the axisymmetric motion and neglecting the hydrodynamic interactions, the non-dimensional swimming translational velocity (scaled with the velocity of isolated phoretic particle) of the swimmer in the presence of the wall can be deduced by applying RRT \cite{Nima_3}
\begin{eqnarray}
U = 1 + \frac{{4{\zeta _1}{\kappa ^3}}}{{3{\Delta ^3}\left( {1 - {{\cos }^2}{\theta _{{\rm{cap}}}}} \right)}},
\end{eqnarray}
Considering that $\zeta_1\sim \Delta^{-2}$ (see Eq.~\ref{zeta_1}), it can be inferred that the far field velocity disturbance created by the obstacle attenuates as $\sim \Delta^{-5}$ and thus the influence of the obstacle is only important in the near field for a phoretic swimmer.

\begin{thebibliography}{64}%
\makeatletter
\providecommand \@ifxundefined [1]{%
 \@ifx{#1\undefined}
}%
\providecommand \@ifnum [1]{%
 \ifnum #1\expandafter \@firstoftwo
 \else \expandafter \@secondoftwo
 \fi
}%
\providecommand \@ifx [1]{%
 \ifx #1\expandafter \@firstoftwo
 \else \expandafter \@secondoftwo
 \fi
}%
\providecommand \natexlab [1]{#1}%
\providecommand \enquote  [1]{``#1''}%
\providecommand \bibnamefont  [1]{#1}%
\providecommand \bibfnamefont [1]{#1}%
\providecommand \citenamefont [1]{#1}%
\providecommand \href@noop [0]{\@secondoftwo}%
\providecommand \href [0]{\begingroup \@sanitize@url \@href}%
\providecommand \@href[1]{\@@startlink{#1}\@@href}%
\providecommand \@@href[1]{\endgroup#1\@@endlink}%
\providecommand \@sanitize@url [0]{\catcode `\\12\catcode `\$12\catcode
  `\&12\catcode `\#12\catcode `\^12\catcode `\_12\catcode `\%12\relax}%
\providecommand \@@startlink[1]{}%
\providecommand \@@endlink[0]{}%
\providecommand \url  [0]{\begingroup\@sanitize@url \@url }%
\providecommand \@url [1]{\endgroup\@href {#1}{\urlprefix }}%
\providecommand \urlprefix  [0]{URL }%
\providecommand \Eprint [0]{\href }%
\providecommand \doibase [0]{http://dx.doi.org/}%
\providecommand \selectlanguage [0]{\@gobble}%
\providecommand \bibinfo  [0]{\@secondoftwo}%
\providecommand \bibfield  [0]{\@secondoftwo}%
\providecommand \translation [1]{[#1]}%
\providecommand \BibitemOpen [0]{}%
\providecommand \bibitemStop [0]{}%
\providecommand \bibitemNoStop [0]{.\EOS\space}%
\providecommand \EOS [0]{\spacefactor3000\relax}%
\providecommand \BibitemShut  [1]{\csname bibitem#1\endcsname}%
\let\auto@bib@innerbib\@empty
\bibitem [{\citenamefont {Lauga}\ \emph {et~al.}(2006)\citenamefont {Lauga},
  \citenamefont {DiLuzio}, \citenamefont {Whitesides},\ and\ \citenamefont
  {Stone}}]{Lauga_2006}%
  \BibitemOpen
  \bibfield  {author} {\bibinfo {author} {\bibfnamefont {E.}~\bibnamefont
  {Lauga}}, \bibinfo {author} {\bibfnamefont {W.}~\bibnamefont {DiLuzio}},
  \bibinfo {author} {\bibfnamefont {G.}~\bibnamefont {Whitesides}}, \ and\
  \bibinfo {author} {\bibfnamefont {H.}~\bibnamefont {Stone}},\ }\href
  {\doibase 10.1529/biophysj.105.069401} {\bibfield  {journal} {\bibinfo
  {journal} {Biophys. J.}\ }\textbf {\bibinfo {volume} {90}},\ \bibinfo {pages}
  {400} (\bibinfo {year} {2006})}\BibitemShut {NoStop}%
\bibitem [{\citenamefont {Di~Leonardo}\ \emph {et~al.}(2010)\citenamefont
  {Di~Leonardo}, \citenamefont {Angelani}, \citenamefont {Dell'Arciprete},
  \citenamefont {Ruocco}, \citenamefont {Iebba}, \citenamefont {Schippa},
  \citenamefont {Conte}, \citenamefont {Mecarini}, \citenamefont {De~Angelis},\
  and\ \citenamefont {Di~Fabrizio}}]{DiLeonardo25052010}%
  \BibitemOpen
  \bibfield  {author} {\bibinfo {author} {\bibfnamefont {R.}~\bibnamefont
  {Di~Leonardo}}, \bibinfo {author} {\bibfnamefont {L.}~\bibnamefont
  {Angelani}}, \bibinfo {author} {\bibfnamefont {D.}~\bibnamefont
  {Dell'Arciprete}}, \bibinfo {author} {\bibfnamefont {G.}~\bibnamefont
  {Ruocco}}, \bibinfo {author} {\bibfnamefont {V.}~\bibnamefont {Iebba}},
  \bibinfo {author} {\bibfnamefont {S.}~\bibnamefont {Schippa}}, \bibinfo
  {author} {\bibfnamefont {M.~P.}\ \bibnamefont {Conte}}, \bibinfo {author}
  {\bibfnamefont {F.}~\bibnamefont {Mecarini}}, \bibinfo {author}
  {\bibfnamefont {F.}~\bibnamefont {De~Angelis}}, \ and\ \bibinfo {author}
  {\bibfnamefont {E.}~\bibnamefont {Di~Fabrizio}},\ }\href {\doibase
  10.1073/pnas.0910426107} {\bibfield  {journal} {\bibinfo  {journal} {Proc.
  Nat. Acad. Sci.}\ }\textbf {\bibinfo {volume} {107}},\ \bibinfo {pages}
  {9541} (\bibinfo {year} {2010})}\BibitemShut {NoStop}%
\bibitem [{\citenamefont {Di~Leonardo}\ \emph {et~al.}(2011)\citenamefont
  {Di~Leonardo}, \citenamefont {Dell'Arciprete}, \citenamefont {Angelani},\
  and\ \citenamefont {Iebba}}]{DiLeonardo2011}%
  \BibitemOpen
  \bibfield  {author} {\bibinfo {author} {\bibfnamefont {R.}~\bibnamefont
  {Di~Leonardo}}, \bibinfo {author} {\bibfnamefont {D.}~\bibnamefont
  {Dell'Arciprete}}, \bibinfo {author} {\bibfnamefont {L.}~\bibnamefont
  {Angelani}}, \ and\ \bibinfo {author} {\bibfnamefont {V.}~\bibnamefont
  {Iebba}},\ }\href {\doibase 10.1103/PhysRevLett.106.038101} {\bibfield
  {journal} {\bibinfo  {journal} {Phys. Rev. Lett.}\ }\textbf {\bibinfo
  {volume} {106}},\ \bibinfo {pages} {038101} (\bibinfo {year}
  {2011})}\BibitemShut {NoStop}%
\bibitem [{\citenamefont {Volpe}\ \emph {et~al.}(2011)\citenamefont {Volpe},
  \citenamefont {Buttinoni}, \citenamefont {Vogt}, \citenamefont {Kummerer},\
  and\ \citenamefont {Bechinger}}]{Volpe}%
  \BibitemOpen
  \bibfield  {author} {\bibinfo {author} {\bibfnamefont {G.}~\bibnamefont
  {Volpe}}, \bibinfo {author} {\bibfnamefont {I.}~\bibnamefont {Buttinoni}},
  \bibinfo {author} {\bibfnamefont {D.}~\bibnamefont {Vogt}}, \bibinfo {author}
  {\bibfnamefont {H.}~\bibnamefont {Kummerer}}, \ and\ \bibinfo {author}
  {\bibfnamefont {C.}~\bibnamefont {Bechinger}},\ }\href {\doibase
  10.1039/C1SM05960B} {\bibfield  {journal} {\bibinfo  {journal} {Soft Matter}\
  }\textbf {\bibinfo {volume} {7}},\ \bibinfo {pages} {8810} (\bibinfo {year}
  {2011})}\BibitemShut {NoStop}%
\bibitem [{\citenamefont {Simmchen}\ \emph {et~al.}(2016)\citenamefont
  {Simmchen}, \citenamefont {Katuri}, \citenamefont {Uspal}, \citenamefont
  {Popescu}, \citenamefont {Tasinkevych},\ and\ \citenamefont
  {S{\'a}nchez}}]{Simmchen_2016}%
  \BibitemOpen
  \bibfield  {author} {\bibinfo {author} {\bibfnamefont {J.}~\bibnamefont
  {Simmchen}}, \bibinfo {author} {\bibfnamefont {J.}~\bibnamefont {Katuri}},
  \bibinfo {author} {\bibfnamefont {W.}~\bibnamefont {Uspal}}, \bibinfo
  {author} {\bibfnamefont {M.}~\bibnamefont {Popescu}}, \bibinfo {author}
  {\bibfnamefont {M.}~\bibnamefont {Tasinkevych}}, \ and\ \bibinfo {author}
  {\bibfnamefont {S.}~\bibnamefont {S{\'a}nchez}},\ }\href {\doibase
  10.1038/ncomms10598} {\bibfield  {journal} {\bibinfo  {journal} {Nat.
  Commun.}\ }\textbf {\bibinfo {volume} {7}},\ \bibinfo {pages} {10598 EP}
  (\bibinfo {year} {2016})}\BibitemShut {NoStop}%
\bibitem [{\citenamefont {Das}\ \emph {et~al.}(2015)\citenamefont {Das},
  \citenamefont {Garg}, \citenamefont {Campbell}, \citenamefont {Howse},
  \citenamefont {Sen}, \citenamefont {Velegol}, \citenamefont {Golestanian},\
  and\ \citenamefont {Ebbens}}]{Velegol_2015}%
  \BibitemOpen
  \bibfield  {author} {\bibinfo {author} {\bibfnamefont {S.}~\bibnamefont
  {Das}}, \bibinfo {author} {\bibfnamefont {A.}~\bibnamefont {Garg}}, \bibinfo
  {author} {\bibfnamefont {A.}~\bibnamefont {Campbell}}, \bibinfo {author}
  {\bibfnamefont {J.}~\bibnamefont {Howse}}, \bibinfo {author} {\bibfnamefont
  {A.}~\bibnamefont {Sen}}, \bibinfo {author} {\bibfnamefont {D.}~\bibnamefont
  {Velegol}}, \bibinfo {author} {\bibfnamefont {R.}~\bibnamefont
  {Golestanian}}, \ and\ \bibinfo {author} {\bibfnamefont {S.}~\bibnamefont
  {Ebbens}},\ }\href {\doibase 10.1038/ncomms9999} {\bibfield  {journal}
  {\bibinfo  {journal} {Nat. Commun.}\ }\textbf {\bibinfo {volume} {6}},\
  \bibinfo {pages} {8999 EP} (\bibinfo {year} {2015})}\BibitemShut {NoStop}%
\bibitem [{\citenamefont {Brown}\ \emph {et~al.}(2016)\citenamefont {Brown},
  \citenamefont {Vladescu}, \citenamefont {Dawson}, \citenamefont {Vissers},
  \citenamefont {Schwarz-Linek}, \citenamefont {Lintuvuori},\ and\
  \citenamefont {Poon}}]{Poon_2016}%
  \BibitemOpen
  \bibfield  {author} {\bibinfo {author} {\bibfnamefont {A.}~\bibnamefont
  {Brown}}, \bibinfo {author} {\bibfnamefont {I.}~\bibnamefont {Vladescu}},
  \bibinfo {author} {\bibfnamefont {A.}~\bibnamefont {Dawson}}, \bibinfo
  {author} {\bibfnamefont {T.}~\bibnamefont {Vissers}}, \bibinfo {author}
  {\bibfnamefont {J.}~\bibnamefont {Schwarz-Linek}}, \bibinfo {author}
  {\bibfnamefont {J.}~\bibnamefont {Lintuvuori}}, \ and\ \bibinfo {author}
  {\bibfnamefont {W.}~\bibnamefont {Poon}},\ }\href {\doibase
  10.1039/C5SM01831E} {\bibfield  {journal} {\bibinfo  {journal} {Soft Matter}\
  }\textbf {\bibinfo {volume} {12}},\ \bibinfo {pages} {131} (\bibinfo {year}
  {2016})}\BibitemShut {NoStop}%
\bibitem [{\citenamefont {O'Toole}\ \emph {et~al.}(2000)\citenamefont
  {O'Toole}, \citenamefont {Kaplan},\ and\ \citenamefont
  {Kolter}}]{Kolter_review}%
  \BibitemOpen
  \bibfield  {author} {\bibinfo {author} {\bibfnamefont {G.}~\bibnamefont
  {O'Toole}}, \bibinfo {author} {\bibfnamefont {H.}~\bibnamefont {Kaplan}}, \
  and\ \bibinfo {author} {\bibfnamefont {R.}~\bibnamefont {Kolter}},\ }\href
  {\doibase 10.1146/annurev.micro.54.1.49} {\bibfield  {journal} {\bibinfo
  {journal} {Ann. Rev. Microbiol.}\ }\textbf {\bibinfo {volume} {54}},\
  \bibinfo {pages} {49} (\bibinfo {year} {2000})}\BibitemShut {NoStop}%
\bibitem [{\citenamefont {van Loosdrecht}\ \emph {et~al.}(1990)\citenamefont
  {van Loosdrecht}, \citenamefont {Lyklema}, \citenamefont {Norde},\ and\
  \citenamefont {Zehnder}}]{Biofilm-rev}%
  \BibitemOpen
  \bibfield  {author} {\bibinfo {author} {\bibfnamefont {M.}~\bibnamefont {van
  Loosdrecht}}, \bibinfo {author} {\bibfnamefont {J.}~\bibnamefont {Lyklema}},
  \bibinfo {author} {\bibfnamefont {W.}~\bibnamefont {Norde}}, \ and\ \bibinfo
  {author} {\bibfnamefont {A.}~\bibnamefont {Zehnder}},\ }\href@noop {}
  {\bibfield  {journal} {\bibinfo  {journal} {Microbiol. Rev.}\ }\textbf
  {\bibinfo {volume} {54}},\ \bibinfo {pages} {75} (\bibinfo {year}
  {1990})}\BibitemShut {NoStop}%
\bibitem [{\citenamefont {Vaccari}\ \emph {et~al.}(2015)\citenamefont
  {Vaccari}, \citenamefont {Allan}, \citenamefont {Sharifi-Mood}, \citenamefont
  {Singh}, \citenamefont {Leheny},\ and\ \citenamefont {Stebe}}]{Vaccari_2015}%
  \BibitemOpen
  \bibfield  {author} {\bibinfo {author} {\bibfnamefont {L.}~\bibnamefont
  {Vaccari}}, \bibinfo {author} {\bibfnamefont {D.}~\bibnamefont {Allan}},
  \bibinfo {author} {\bibfnamefont {N.}~\bibnamefont {Sharifi-Mood}}, \bibinfo
  {author} {\bibfnamefont {A.}~\bibnamefont {Singh}}, \bibinfo {author}
  {\bibfnamefont {R.}~\bibnamefont {Leheny}}, \ and\ \bibinfo {author}
  {\bibfnamefont {K.}~\bibnamefont {Stebe}},\ }\href {\doibase
  10.1039/C5SM00696A} {\bibfield  {journal} {\bibinfo  {journal} {Soft Matter}\
  }\textbf {\bibinfo {volume} {11}},\ \bibinfo {pages} {6062} (\bibinfo {year}
  {2015})}\BibitemShut {NoStop}%
\bibitem [{\citenamefont {Berke}\ \emph {et~al.}(2008)\citenamefont {Berke},
  \citenamefont {Turner}, \citenamefont {Berg},\ and\ \citenamefont
  {Lauga}}]{Lauga_2008}%
  \BibitemOpen
  \bibfield  {author} {\bibinfo {author} {\bibfnamefont {A.}~\bibnamefont
  {Berke}}, \bibinfo {author} {\bibfnamefont {L.}~\bibnamefont {Turner}},
  \bibinfo {author} {\bibfnamefont {H.}~\bibnamefont {Berg}}, \ and\ \bibinfo
  {author} {\bibfnamefont {E.}~\bibnamefont {Lauga}},\ }\href {\doibase
  10.1103/PhysRevLett.101.038102} {\bibfield  {journal} {\bibinfo  {journal}
  {Phys. Rev. Lett.}\ }\textbf {\bibinfo {volume} {101}},\ \bibinfo {pages}
  {038102} (\bibinfo {year} {2008})}\BibitemShut {NoStop}%
\bibitem [{\citenamefont {Spagnolie}\ and\ \citenamefont
  {Lauga}(2012)}]{spagnolie_lauga_2012}%
  \BibitemOpen
  \bibfield  {author} {\bibinfo {author} {\bibfnamefont {S.}~\bibnamefont
  {Spagnolie}}\ and\ \bibinfo {author} {\bibfnamefont {E.}~\bibnamefont
  {Lauga}},\ }\href {\doibase 10.1017/jfm.2012.101} {\bibfield  {journal}
  {\bibinfo  {journal} {J. Fluid Mech.}\ }\textbf {\bibinfo {volume} {700}},\
  \bibinfo {pages} {105} (\bibinfo {year} {2012})}\BibitemShut {NoStop}%
\bibitem [{\citenamefont {Lighthill}(1952)}]{Lighthill}%
  \BibitemOpen
  \bibfield  {author} {\bibinfo {author} {\bibfnamefont {M.~J.}\ \bibnamefont
  {Lighthill}},\ }\href {\doibase 10.1002/cpa.3160050201} {\bibfield  {journal}
  {\bibinfo  {journal} {Comm. Pure Appl. Math.}\ }\textbf {\bibinfo {volume}
  {5}},\ \bibinfo {pages} {109} (\bibinfo {year} {1952})}\BibitemShut {NoStop}%
\bibitem [{\citenamefont {Najafi}\ and\ \citenamefont
  {Golestanian}(2004)}]{Najafi}%
  \BibitemOpen
  \bibfield  {author} {\bibinfo {author} {\bibfnamefont {A.}~\bibnamefont
  {Najafi}}\ and\ \bibinfo {author} {\bibfnamefont {R.}~\bibnamefont
  {Golestanian}},\ }\href {\doibase 10.1103/PhysRevE.69.062901} {\bibfield
  {journal} {\bibinfo  {journal} {Phys. Rev. E}\ }\textbf {\bibinfo {volume}
  {69}},\ \bibinfo {pages} {062901} (\bibinfo {year} {2004})}\BibitemShut
  {NoStop}%
\bibitem [{\citenamefont {Lauga}\ and\ \citenamefont
  {Powers}(2009)}]{Lauga_Powers}%
  \BibitemOpen
  \bibfield  {author} {\bibinfo {author} {\bibfnamefont {E.}~\bibnamefont
  {Lauga}}\ and\ \bibinfo {author} {\bibfnamefont {T.}~\bibnamefont {Powers}},\
  }\href {\doibase 0034-4885/72/i=9/a=096601} {\bibfield  {journal} {\bibinfo
  {journal} {Rep. Prog. Phys.}\ }\textbf {\bibinfo {volume} {72}},\ \bibinfo
  {pages} {096601} (\bibinfo {year} {2009})}\BibitemShut {NoStop}%
\bibitem [{\citenamefont {Spagnolie}\ \emph {et~al.}(2015)\citenamefont
  {Spagnolie}, \citenamefont {Moreno-Flores}, \citenamefont {Bartolo},\ and\
  \citenamefont {Lauga}}]{Spagnolie_Lauga_2015}%
  \BibitemOpen
  \bibfield  {author} {\bibinfo {author} {\bibfnamefont {S.}~\bibnamefont
  {Spagnolie}}, \bibinfo {author} {\bibfnamefont {G.}~\bibnamefont
  {Moreno-Flores}}, \bibinfo {author} {\bibfnamefont {D.}~\bibnamefont
  {Bartolo}}, \ and\ \bibinfo {author} {\bibfnamefont {E.}~\bibnamefont
  {Lauga}},\ }\href {\doibase 10.1039/C4SM02785J} {\bibfield  {journal}
  {\bibinfo  {journal} {Soft Matter}\ }\textbf {\bibinfo {volume} {11}},\
  \bibinfo {pages} {3396} (\bibinfo {year} {2015})}\BibitemShut {NoStop}%
\bibitem [{\citenamefont {Jiang}\ \emph {et~al.}(2010)\citenamefont {Jiang},
  \citenamefont {Yoshinaga},\ and\ \citenamefont {Sano}}]{Jiang}%
  \BibitemOpen
  \bibfield  {author} {\bibinfo {author} {\bibfnamefont {H.-R.}\ \bibnamefont
  {Jiang}}, \bibinfo {author} {\bibfnamefont {N.}~\bibnamefont {Yoshinaga}}, \
  and\ \bibinfo {author} {\bibfnamefont {M.}~\bibnamefont {Sano}},\ }\href
  {\doibase 10.1103/PhysRevLett.105.268302} {\bibfield  {journal} {\bibinfo
  {journal} {Phys. Rev. Lett.}\ }\textbf {\bibinfo {volume} {105}},\ \bibinfo
  {pages} {268302} (\bibinfo {year} {2010})}\BibitemShut {NoStop}%
\bibitem [{\citenamefont {Paxton}\ \emph {et~al.}(2004)\citenamefont {Paxton},
  \citenamefont {Kistler}, \citenamefont {Olmeda}, \citenamefont {Sen},
  \citenamefont {St.~Angelo}, \citenamefont {Cao}, \citenamefont {Mallouk},
  \citenamefont {Lammert},\ and\ \citenamefont {Crespi}}]{Sen2004}%
  \BibitemOpen
  \bibfield  {author} {\bibinfo {author} {\bibfnamefont {W.}~\bibnamefont
  {Paxton}}, \bibinfo {author} {\bibfnamefont {K.}~\bibnamefont {Kistler}},
  \bibinfo {author} {\bibfnamefont {C.}~\bibnamefont {Olmeda}}, \bibinfo
  {author} {\bibfnamefont {A.}~\bibnamefont {Sen}}, \bibinfo {author}
  {\bibfnamefont {S.}~\bibnamefont {St.~Angelo}}, \bibinfo {author}
  {\bibfnamefont {Y.}~\bibnamefont {Cao}}, \bibinfo {author} {\bibfnamefont
  {T.}~\bibnamefont {Mallouk}}, \bibinfo {author} {\bibfnamefont
  {P.}~\bibnamefont {Lammert}}, \ and\ \bibinfo {author} {\bibfnamefont
  {V.}~\bibnamefont {Crespi}},\ }\href {\doibase 10.1021/ja047697z} {\bibfield
  {journal} {\bibinfo  {journal} {J. Am. Chem. Soc.}\ }\textbf {\bibinfo
  {volume} {126}},\ \bibinfo {pages} {13424} (\bibinfo {year}
  {2004})}\BibitemShut {NoStop}%
\bibitem [{\citenamefont {Howse}\ \emph {et~al.}(2007)\citenamefont {Howse},
  \citenamefont {Jones}, \citenamefont {Ryan}, \citenamefont {Gough},
  \citenamefont {Vafabakhsh},\ and\ \citenamefont {Golestanian}}]{Howse2007}%
  \BibitemOpen
  \bibfield  {author} {\bibinfo {author} {\bibfnamefont {J.~R.}\ \bibnamefont
  {Howse}}, \bibinfo {author} {\bibfnamefont {R.~A.~L.}\ \bibnamefont {Jones}},
  \bibinfo {author} {\bibfnamefont {A.~J.}\ \bibnamefont {Ryan}}, \bibinfo
  {author} {\bibfnamefont {T.}~\bibnamefont {Gough}}, \bibinfo {author}
  {\bibfnamefont {R.}~\bibnamefont {Vafabakhsh}}, \ and\ \bibinfo {author}
  {\bibfnamefont {R.}~\bibnamefont {Golestanian}},\ }\href {\doibase
  10.1103/PhysRevLett.99.048102} {\bibfield  {journal} {\bibinfo  {journal}
  {Phys. Rev. Lett.}\ }\textbf {\bibinfo {volume} {99}},\ \bibinfo {pages}
  {048102} (\bibinfo {year} {2007})}\BibitemShut {NoStop}%
\bibitem [{\citenamefont {Sanchez}\ \emph {et~al.}(2011)\citenamefont
  {Sanchez}, \citenamefont {Solovev}, \citenamefont {Schulze},\ and\
  \citenamefont {Schmidt}}]{Schmidt2001}%
  \BibitemOpen
  \bibfield  {author} {\bibinfo {author} {\bibfnamefont {A.}~\bibnamefont
  {Sanchez}}, \bibinfo {author} {\bibfnamefont {A.}~\bibnamefont {Solovev}},
  \bibinfo {author} {\bibfnamefont {S.}~\bibnamefont {Schulze}}, \ and\
  \bibinfo {author} {\bibfnamefont {O.}~\bibnamefont {Schmidt}},\ }\href
  {\doibase 10.1039/C0CC04126B} {\bibfield  {journal} {\bibinfo  {journal}
  {Chem. Commun.}\ }\textbf {\bibinfo {volume} {47}},\ \bibinfo {pages} {698}
  (\bibinfo {year} {2011})}\BibitemShut {NoStop}%
\bibitem [{\citenamefont {Patra}\ \emph {et~al.}(2013)\citenamefont {Patra},
  \citenamefont {Sengupta}, \citenamefont {Duan}, \citenamefont {Zhang},
  \citenamefont {Pavlick},\ and\ \citenamefont {Sen}}]{Patra_2013}%
  \BibitemOpen
  \bibfield  {author} {\bibinfo {author} {\bibfnamefont {D.}~\bibnamefont
  {Patra}}, \bibinfo {author} {\bibfnamefont {S.}~\bibnamefont {Sengupta}},
  \bibinfo {author} {\bibfnamefont {W.}~\bibnamefont {Duan}}, \bibinfo {author}
  {\bibfnamefont {H.}~\bibnamefont {Zhang}}, \bibinfo {author} {\bibfnamefont
  {R.}~\bibnamefont {Pavlick}}, \ and\ \bibinfo {author} {\bibfnamefont
  {A.}~\bibnamefont {Sen}},\ }\href {\doibase 10.1039/C2NR32600K} {\bibfield
  {journal} {\bibinfo  {journal} {Nanoscale}\ }\textbf {\bibinfo {volume}
  {5}},\ \bibinfo {pages} {1273} (\bibinfo {year} {2013})}\BibitemShut
  {NoStop}%
\bibitem [{\citenamefont {Gao}\ and\ \citenamefont
  {Wang}(2014{\natexlab{a}})}]{Wang_2014}%
  \BibitemOpen
  \bibfield  {author} {\bibinfo {author} {\bibfnamefont {W.}~\bibnamefont
  {Gao}}\ and\ \bibinfo {author} {\bibfnamefont {J.}~\bibnamefont {Wang}},\
  }\href {\doibase 10.1039/C4NR03124E} {\bibfield  {journal} {\bibinfo
  {journal} {Nanoscale}\ }\textbf {\bibinfo {volume} {6}},\ \bibinfo {pages}
  {10486} (\bibinfo {year} {2014}{\natexlab{a}})}\BibitemShut {NoStop}%
\bibitem [{\citenamefont {Gao}\ and\ \citenamefont
  {Wang}(2014{\natexlab{b}})}]{Wang_Pul1}%
  \BibitemOpen
  \bibfield  {author} {\bibinfo {author} {\bibfnamefont {W.}~\bibnamefont
  {Gao}}\ and\ \bibinfo {author} {\bibfnamefont {J.}~\bibnamefont {Wang}},\
  }\href {\doibase 10.1021/nn500077a} {\bibfield  {journal} {\bibinfo
  {journal} {ACS Nano}\ }\textbf {\bibinfo {volume} {8}},\ \bibinfo {pages}
  {3170} (\bibinfo {year} {2014}{\natexlab{b}})}\BibitemShut {NoStop}%
\bibitem [{\citenamefont {Burdick}\ \emph {et~al.}(2008)\citenamefont
  {Burdick}, \citenamefont {Laocharoensuk}, \citenamefont {Wheat},
  \citenamefont {Posner},\ and\ \citenamefont {Wang}}]{Burdick_2008}%
  \BibitemOpen
  \bibfield  {author} {\bibinfo {author} {\bibfnamefont {J.}~\bibnamefont
  {Burdick}}, \bibinfo {author} {\bibfnamefont {R.}~\bibnamefont
  {Laocharoensuk}}, \bibinfo {author} {\bibfnamefont {P.~M.}\ \bibnamefont
  {Wheat}}, \bibinfo {author} {\bibfnamefont {J.~D.}\ \bibnamefont {Posner}}, \
  and\ \bibinfo {author} {\bibfnamefont {J.}~\bibnamefont {Wang}},\ }\href
  {\doibase 10.1021/ja803529u} {\bibfield  {journal} {\bibinfo  {journal} {J.
  Am. Chem. Soc.}\ }\textbf {\bibinfo {volume} {130}},\ \bibinfo {pages} {8164}
  (\bibinfo {year} {2008})}\BibitemShut {NoStop}%
\bibitem [{\citenamefont {Gao}\ \emph {et~al.}(2012)\citenamefont {Gao},
  \citenamefont {Kagan}, \citenamefont {Pak}, \citenamefont {Clawson},
  \citenamefont {Campuzano}, \citenamefont {Chuluun-Erdene}, \citenamefont
  {Shipton}, \citenamefont {Fullerton}, \citenamefont {Zhang}, \citenamefont
  {Lauga},\ and\ \citenamefont {Wang}}]{Wang_Magnet}%
  \BibitemOpen
  \bibfield  {author} {\bibinfo {author} {\bibfnamefont {W.}~\bibnamefont
  {Gao}}, \bibinfo {author} {\bibfnamefont {D.}~\bibnamefont {Kagan}}, \bibinfo
  {author} {\bibfnamefont {O.~S.}\ \bibnamefont {Pak}}, \bibinfo {author}
  {\bibfnamefont {C.}~\bibnamefont {Clawson}}, \bibinfo {author} {\bibfnamefont
  {S.}~\bibnamefont {Campuzano}}, \bibinfo {author} {\bibfnamefont
  {E.}~\bibnamefont {Chuluun-Erdene}}, \bibinfo {author} {\bibfnamefont
  {E.}~\bibnamefont {Shipton}}, \bibinfo {author} {\bibfnamefont {E.~E.}\
  \bibnamefont {Fullerton}}, \bibinfo {author} {\bibfnamefont {L.}~\bibnamefont
  {Zhang}}, \bibinfo {author} {\bibfnamefont {E.}~\bibnamefont {Lauga}}, \ and\
  \bibinfo {author} {\bibfnamefont {J.}~\bibnamefont {Wang}},\ }\href {\doibase
  10.1002/smll.201101909} {\bibfield  {journal} {\bibinfo  {journal} {Small}\
  }\textbf {\bibinfo {volume} {8}},\ \bibinfo {pages} {460} (\bibinfo {year}
  {2012})}\BibitemShut {NoStop}%
\bibitem [{\citenamefont {Campbell}\ and\ \citenamefont
  {Ebbens}(2013)}]{Campbell_2013}%
  \BibitemOpen
  \bibfield  {author} {\bibinfo {author} {\bibfnamefont {A.}~\bibnamefont
  {Campbell}}\ and\ \bibinfo {author} {\bibfnamefont {S.}~\bibnamefont
  {Ebbens}},\ }\href {\doibase 10.1021/la403450j} {\bibfield  {journal}
  {\bibinfo  {journal} {Langmuir}\ }\textbf {\bibinfo {volume} {29}},\ \bibinfo
  {pages} {14066} (\bibinfo {year} {2013})}\BibitemShut {NoStop}%
\bibitem [{\citenamefont {ten Hagen}\ \emph {et~al.}(2014)\citenamefont {ten
  Hagen}, \citenamefont {K{\"u}mmel}, \citenamefont {Wittkowski}, \citenamefont
  {Takagi}, \citenamefont {L{\"o}wen},\ and\ \citenamefont
  {Bechinger}}]{Bechinger_2014}%
  \BibitemOpen
  \bibfield  {author} {\bibinfo {author} {\bibfnamefont {B.}~\bibnamefont {ten
  Hagen}}, \bibinfo {author} {\bibfnamefont {F.}~\bibnamefont {K{\"u}mmel}},
  \bibinfo {author} {\bibfnamefont {R.}~\bibnamefont {Wittkowski}}, \bibinfo
  {author} {\bibfnamefont {D.}~\bibnamefont {Takagi}}, \bibinfo {author}
  {\bibfnamefont {H.}~\bibnamefont {L{\"o}wen}}, \ and\ \bibinfo {author}
  {\bibfnamefont {C.}~\bibnamefont {Bechinger}},\ }\href {\doibase
  10.1038/ncomms5829} {\bibfield  {journal} {\bibinfo  {journal} {Nat.
  Commun.}\ }\textbf {\bibinfo {volume} {5}},\ \bibinfo {pages} {4829 EP}
  (\bibinfo {year} {2014})}\BibitemShut {NoStop}%
\bibitem [{\citenamefont {Hong}\ \emph {et~al.}(2007)\citenamefont {Hong},
  \citenamefont {Blackman}, \citenamefont {Kopp}, \citenamefont {Sen},\ and\
  \citenamefont {Velegol}}]{Velegol_2007}%
  \BibitemOpen
  \bibfield  {author} {\bibinfo {author} {\bibfnamefont {Y.}~\bibnamefont
  {Hong}}, \bibinfo {author} {\bibfnamefont {N.~M.~K.}\ \bibnamefont
  {Blackman}}, \bibinfo {author} {\bibfnamefont {N.~D.}\ \bibnamefont {Kopp}},
  \bibinfo {author} {\bibfnamefont {A.}~\bibnamefont {Sen}}, \ and\ \bibinfo
  {author} {\bibfnamefont {D.}~\bibnamefont {Velegol}},\ }\href {\doibase
  10.1103/PhysRevLett.99.178103} {\bibfield  {journal} {\bibinfo  {journal}
  {Phys. Rev. Lett.}\ }\textbf {\bibinfo {volume} {99}},\ \bibinfo {pages}
  {178103} (\bibinfo {year} {2007})}\BibitemShut {NoStop}%
\bibitem [{\citenamefont {Baraban}\ \emph {et~al.}(2013)\citenamefont
  {Baraban}, \citenamefont {Harazim}, \citenamefont {Sanchez},\ and\
  \citenamefont {Schmidt}}]{Baraban_2013}%
  \BibitemOpen
  \bibfield  {author} {\bibinfo {author} {\bibfnamefont {L.}~\bibnamefont
  {Baraban}}, \bibinfo {author} {\bibfnamefont {S.}~\bibnamefont {Harazim}},
  \bibinfo {author} {\bibfnamefont {S.}~\bibnamefont {Sanchez}}, \ and\
  \bibinfo {author} {\bibfnamefont {O.}~\bibnamefont {Schmidt}},\ }\href
  {\doibase 10.1002/anie.201301460} {\bibfield  {journal} {\bibinfo  {journal}
  {Angew. Chem. Int. Ed.}\ }\textbf {\bibinfo {volume} {52}},\ \bibinfo {pages}
  {5552} (\bibinfo {year} {2013})}\BibitemShut {NoStop}%
\bibitem [{\citenamefont {Takagi}\ \emph {et~al.}(2014)\citenamefont {Takagi},
  \citenamefont {Palacci}, \citenamefont {Braunschweig}, \citenamefont
  {Shelley},\ and\ \citenamefont {Zhang}}]{Takagi_2014}%
  \BibitemOpen
  \bibfield  {author} {\bibinfo {author} {\bibfnamefont {D.}~\bibnamefont
  {Takagi}}, \bibinfo {author} {\bibfnamefont {J.}~\bibnamefont {Palacci}},
  \bibinfo {author} {\bibfnamefont {A.}~\bibnamefont {Braunschweig}}, \bibinfo
  {author} {\bibfnamefont {M.}~\bibnamefont {Shelley}}, \ and\ \bibinfo
  {author} {\bibfnamefont {J.}~\bibnamefont {Zhang}},\ }\href {\doibase
  10.1039/C3SM52815D} {\bibfield  {journal} {\bibinfo  {journal} {Soft Matter}\
  }\textbf {\bibinfo {volume} {10}},\ \bibinfo {pages} {1784} (\bibinfo {year}
  {2014})}\BibitemShut {NoStop}%
\bibitem [{\citenamefont {Davies~Wykes}\ \emph {et~al.}(2017)\citenamefont
  {Davies~Wykes}, \citenamefont {Zhong}, \citenamefont {Tong}, \citenamefont
  {Adachi}, \citenamefont {Liu}, \citenamefont {Ristroph}, \citenamefont
  {Ward}, \citenamefont {Shelley},\ and\ \citenamefont {Zhang}}]{Wykes_2017}%
  \BibitemOpen
  \bibfield  {author} {\bibinfo {author} {\bibfnamefont {M.}~\bibnamefont
  {Davies~Wykes}}, \bibinfo {author} {\bibfnamefont {X.}~\bibnamefont {Zhong}},
  \bibinfo {author} {\bibfnamefont {J.}~\bibnamefont {Tong}}, \bibinfo {author}
  {\bibfnamefont {T.}~\bibnamefont {Adachi}}, \bibinfo {author} {\bibfnamefont
  {Y.}~\bibnamefont {Liu}}, \bibinfo {author} {\bibfnamefont {L.}~\bibnamefont
  {Ristroph}}, \bibinfo {author} {\bibfnamefont {M.}~\bibnamefont {Ward}},
  \bibinfo {author} {\bibfnamefont {M.}~\bibnamefont {Shelley}}, \ and\
  \bibinfo {author} {\bibfnamefont {J.}~\bibnamefont {Zhang}},\ }\href
  {\doibase 10.1039/C7SM00203C} {\bibfield  {journal} {\bibinfo  {journal}
  {Soft Matter}\ }\textbf {\bibinfo {volume} {13}},\ \bibinfo {pages} {4681}
  (\bibinfo {year} {2017})}\BibitemShut {NoStop}%
\bibitem [{\citenamefont {Kreuter}\ \emph {et~al.}(2013)\citenamefont
  {Kreuter}, \citenamefont {Siems}, \citenamefont {Nielaba}, \citenamefont
  {Leiderer},\ and\ \citenamefont {Erbe}}]{Kreuter_2013}%
  \BibitemOpen
  \bibfield  {author} {\bibinfo {author} {\bibfnamefont {C.}~\bibnamefont
  {Kreuter}}, \bibinfo {author} {\bibfnamefont {U.}~\bibnamefont {Siems}},
  \bibinfo {author} {\bibfnamefont {P.}~\bibnamefont {Nielaba}}, \bibinfo
  {author} {\bibfnamefont {P.}~\bibnamefont {Leiderer}}, \ and\ \bibinfo
  {author} {\bibfnamefont {A.}~\bibnamefont {Erbe}},\ }\href {\doibase
  10.1140/epjst/e2013-02067-x} {\bibfield  {journal} {\bibinfo  {journal} {Eur.
  Phys. J. ST}\ }\textbf {\bibinfo {volume} {222}},\ \bibinfo {pages} {2923}
  (\bibinfo {year} {2013})}\BibitemShut {NoStop}%
\bibitem [{\citenamefont {Parmar}\ \emph {et~al.}(2015)\citenamefont {Parmar},
  \citenamefont {Ma}, \citenamefont {Katuri}, \citenamefont {Simmchen},
  \citenamefont {Stanton}, \citenamefont {Trichet-Paredes}, \citenamefont
  {Soler},\ and\ \citenamefont {Sanchez}}]{Parmar_2015}%
  \BibitemOpen
  \bibfield  {author} {\bibinfo {author} {\bibfnamefont {J.}~\bibnamefont
  {Parmar}}, \bibinfo {author} {\bibfnamefont {X.}~\bibnamefont {Ma}}, \bibinfo
  {author} {\bibfnamefont {J.}~\bibnamefont {Katuri}}, \bibinfo {author}
  {\bibfnamefont {J.}~\bibnamefont {Simmchen}}, \bibinfo {author}
  {\bibfnamefont {M.}~\bibnamefont {Stanton}}, \bibinfo {author} {\bibfnamefont
  {C.}~\bibnamefont {Trichet-Paredes}}, \bibinfo {author} {\bibfnamefont
  {L.}~\bibnamefont {Soler}}, \ and\ \bibinfo {author} {\bibfnamefont
  {S.}~\bibnamefont {Sanchez}},\ }\href {\doibase
  10.1088/1468-6996/16/1/014802} {\bibfield  {journal} {\bibinfo  {journal}
  {Sci. Technol. Adv. Mater.}\ }\textbf {\bibinfo {volume} {16}},\ \bibinfo
  {pages} {014802} (\bibinfo {year} {2015})}\BibitemShut {NoStop}%
\bibitem [{\citenamefont {Crowdy}(2013)}]{crowdy_2013}%
  \BibitemOpen
  \bibfield  {author} {\bibinfo {author} {\bibfnamefont {D.}~\bibnamefont
  {Crowdy}},\ }\href {\doibase 10.1017/jfm.2013.510} {\bibfield  {journal}
  {\bibinfo  {journal} {J. Fluid Mech.}\ }\textbf {\bibinfo {volume} {735}},\
  \bibinfo {pages} {473} (\bibinfo {year} {2013})}\BibitemShut {NoStop}%
\bibitem [{\citenamefont {Uspal}\ \emph {et~al.}(2015)\citenamefont {Uspal},
  \citenamefont {Popescu}, \citenamefont {Dietrich},\ and\ \citenamefont
  {Tasinkevych}}]{Uspal}%
  \BibitemOpen
  \bibfield  {author} {\bibinfo {author} {\bibfnamefont {W.~E.}\ \bibnamefont
  {Uspal}}, \bibinfo {author} {\bibfnamefont {M.~N.}\ \bibnamefont {Popescu}},
  \bibinfo {author} {\bibfnamefont {S.}~\bibnamefont {Dietrich}}, \ and\
  \bibinfo {author} {\bibfnamefont {M.}~\bibnamefont {Tasinkevych}},\ }\href
  {\doibase 10.1039/C4SM02317J} {\bibfield  {journal} {\bibinfo  {journal}
  {Soft Matter}\ }\textbf {\bibinfo {volume} {11}},\ \bibinfo {pages} {434}
  (\bibinfo {year} {2015})}\BibitemShut {NoStop}%
\bibitem [{\citenamefont {Ibrahim}\ and\ \citenamefont
  {Liverpool}(2015)}]{Liverpool_2015}%
  \BibitemOpen
  \bibfield  {author} {\bibinfo {author} {\bibfnamefont {Y.}~\bibnamefont
  {Ibrahim}}\ and\ \bibinfo {author} {\bibfnamefont {T.~B.}\ \bibnamefont
  {Liverpool}},\ }\href {\doibase 0295-5075/111/i=4/a=48008} {\bibfield
  {journal} {\bibinfo  {journal} {Eur. Phys. Lett.}\ }\textbf {\bibinfo
  {volume} {111}},\ \bibinfo {pages} {48008} (\bibinfo {year}
  {2015})}\BibitemShut {NoStop}%
\bibitem [{\citenamefont {Ibrahim}\ and\ \citenamefont
  {Liverpool}(2016)}]{Liverpool_2016}%
  \BibitemOpen
  \bibfield  {author} {\bibinfo {author} {\bibfnamefont {Y.}~\bibnamefont
  {Ibrahim}}\ and\ \bibinfo {author} {\bibfnamefont {T.}~\bibnamefont
  {Liverpool}},\ }\href {\doibase 10.1140/epjst/e2016-60148-1} {\bibfield
  {journal} {\bibinfo  {journal} {Eur. Phys. J. ST}\ }\textbf {\bibinfo
  {volume} {225}},\ \bibinfo {pages} {1843} (\bibinfo {year}
  {2016})}\BibitemShut {NoStop}%
\bibitem [{\citenamefont {Mozaffari}\ \emph
  {et~al.}(2016{\natexlab{a}})\citenamefont {Mozaffari}, \citenamefont
  {Sharifi-Mood}, \citenamefont {Koplik},\ and\ \citenamefont
  {Maldarelli}}]{Wall_1}%
  \BibitemOpen
  \bibfield  {author} {\bibinfo {author} {\bibfnamefont {A.}~\bibnamefont
  {Mozaffari}}, \bibinfo {author} {\bibfnamefont {N.}~\bibnamefont
  {Sharifi-Mood}}, \bibinfo {author} {\bibfnamefont {J.}~\bibnamefont
  {Koplik}}, \ and\ \bibinfo {author} {\bibfnamefont {C.}~\bibnamefont
  {Maldarelli}},\ }\href {\doibase 10.1063/1.4948398} {\bibfield  {journal}
  {\bibinfo  {journal} {Phys. Fluids}\ }\textbf {\bibinfo {volume} {28}},\
  \bibinfo {pages} {053107} (\bibinfo {year} {2016}{\natexlab{a}})}\BibitemShut
  {NoStop}%
\bibitem [{\citenamefont {Yariv}(2016)}]{Yariv_2016}%
  \BibitemOpen
  \bibfield  {author} {\bibinfo {author} {\bibfnamefont {E.}~\bibnamefont
  {Yariv}},\ }\href {\doibase 10.1103/PhysRevFluids.1.032101} {\bibfield
  {journal} {\bibinfo  {journal} {Phys. Rev. Fluids}\ }\textbf {\bibinfo
  {volume} {1}},\ \bibinfo {pages} {032101} (\bibinfo {year}
  {2016})}\BibitemShut {NoStop}%
\bibitem [{\citenamefont {Mozaffari}\ \emph
  {et~al.}(2016{\natexlab{b}})\citenamefont {Mozaffari}, \citenamefont
  {Sharifi-Mood}, \citenamefont {Koplik},\ and\ \citenamefont
  {Maldarelli}}]{Wall_2}%
  \BibitemOpen
  \bibfield  {author} {\bibinfo {author} {\bibfnamefont {A.}~\bibnamefont
  {Mozaffari}}, \bibinfo {author} {\bibfnamefont {N.}~\bibnamefont
  {Sharifi-Mood}}, \bibinfo {author} {\bibfnamefont {J.}~\bibnamefont
  {Koplik}}, \ and\ \bibinfo {author} {\bibfnamefont {C.}~\bibnamefont
  {Maldarelli}},\ }\href {\doibase arXiv:1611.03883} {\bibfield  {journal}
  {\bibinfo  {journal} {ArXiv}\ ,\ } (\bibinfo {year}
  {2016}{\natexlab{b}})}\BibitemShut {NoStop}%
\bibitem [{\citenamefont {Papavassiliou}\ and\ \citenamefont
  {Alexander}(2017)}]{Alexander_2017}%
  \BibitemOpen
  \bibfield  {author} {\bibinfo {author} {\bibfnamefont {D.}~\bibnamefont
  {Papavassiliou}}\ and\ \bibinfo {author} {\bibfnamefont {G.}~\bibnamefont
  {Alexander}},\ }\href {\doibase 10.1017/jfm.2016.837} {\bibfield  {journal}
  {\bibinfo  {journal} {J. Fluid Mech.}\ }\textbf {\bibinfo {volume} {813}},\
  \bibinfo {pages} {618} (\bibinfo {year} {2017})}\BibitemShut {NoStop}%
\bibitem [{\citenamefont {Purcell}(1977)}]{Purcell}%
  \BibitemOpen
  \bibfield  {author} {\bibinfo {author} {\bibfnamefont {E.}~\bibnamefont
  {Purcell}},\ }\href {\doibase 10.1119/1.10903} {\bibfield  {journal}
  {\bibinfo  {journal} {Am. J. Phys.}\ }\textbf {\bibinfo {volume} {45}},\
  \bibinfo {pages} {3} (\bibinfo {year} {1977})}\BibitemShut {NoStop}%
\bibitem [{\citenamefont {Lee}\ and\ \citenamefont {Leal}(1980)}]{JFM2}%
  \BibitemOpen
  \bibfield  {author} {\bibinfo {author} {\bibfnamefont {S.~H.}\ \bibnamefont
  {Lee}}\ and\ \bibinfo {author} {\bibfnamefont {L.~G.}\ \bibnamefont {Leal}},\
  }\href {\doibase 10.1017/S0022112080000109} {\bibfield  {journal} {\bibinfo
  {journal} {J. Fluid Mech.}\ }\textbf {\bibinfo {volume} {98}},\ \bibinfo
  {pages} {193} (\bibinfo {year} {1980})}\BibitemShut {NoStop}%
\bibitem [{\citenamefont {Teubner}(1982)}]{teubner}%
  \BibitemOpen
  \bibfield  {author} {\bibinfo {author} {\bibfnamefont {M.}~\bibnamefont
  {Teubner}},\ }\href {\doibase 10.1063/1.442861} {\bibfield  {journal}
  {\bibinfo  {journal} {J. Chem. Phys.}\ }\textbf {\bibinfo {volume} {76}},\
  \bibinfo {pages} {5564} (\bibinfo {year} {1982})}\BibitemShut {NoStop}%
\bibitem [{\citenamefont {Stone}\ and\ \citenamefont
  {Samuel}(1996)}]{Stone_Samuel}%
  \BibitemOpen
  \bibfield  {author} {\bibinfo {author} {\bibfnamefont {H.}~\bibnamefont
  {Stone}}\ and\ \bibinfo {author} {\bibfnamefont {A.}~\bibnamefont {Samuel}},\
  }\href {\doibase 10.1103/PhysRevLett.77.4102} {\bibfield  {journal} {\bibinfo
   {journal} {Phys. Rev. Lett.}\ }\textbf {\bibinfo {volume} {77}},\ \bibinfo
  {pages} {4102} (\bibinfo {year} {1996})}\BibitemShut {NoStop}%
\bibitem [{\citenamefont {Masoud}\ and\ \citenamefont {Stone}(2014)}]{Masood}%
  \BibitemOpen
  \bibfield  {author} {\bibinfo {author} {\bibfnamefont {H.}~\bibnamefont
  {Masoud}}\ and\ \bibinfo {author} {\bibfnamefont {H.~A.}\ \bibnamefont
  {Stone}},\ }\href {\doibase 10.1017/jfm.2014.8} {\bibfield  {journal}
  {\bibinfo  {journal} {J. Fluid Mech.}\ }\textbf {\bibinfo {volume} {741}},\
  \bibinfo {pages} {1} (\bibinfo {year} {2014})}\BibitemShut {NoStop}%
\bibitem [{\citenamefont {Elfring}(2015)}]{Elfring_RRT}%
  \BibitemOpen
  \bibfield  {author} {\bibinfo {author} {\bibfnamefont {G.}~\bibnamefont
  {Elfring}},\ }\href {\doibase http://dx.doi.org/10.1063/1.4906993} {\bibfield
   {journal} {\bibinfo  {journal} {Phys. Fluids}\ }\textbf {\bibinfo {volume}
  {27}},\ \bibinfo {pages} {023101} (\bibinfo {year} {2015})}\BibitemShut
  {NoStop}%
\bibitem [{\citenamefont {Blake}(1971)}]{Blake_1971}%
  \BibitemOpen
  \bibfield  {author} {\bibinfo {author} {\bibfnamefont {J.~R.}\ \bibnamefont
  {Blake}},\ }\href {\doibase 10.1017/S002211207100048X} {\bibfield  {journal}
  {\bibinfo  {journal} {J. Fluid Mech.}\ }\textbf {\bibinfo {volume} {46}},\
  \bibinfo {pages} {199} (\bibinfo {year} {1971})}\BibitemShut {NoStop}%
\bibitem [{\citenamefont {Pak}\ and\ \citenamefont
  {Lauga}(2014)}]{Lauga_Pak_2014}%
  \BibitemOpen
  \bibfield  {author} {\bibinfo {author} {\bibfnamefont {O.}~\bibnamefont
  {Pak}}\ and\ \bibinfo {author} {\bibfnamefont {E.}~\bibnamefont {Lauga}},\
  }\href {\doibase 10.1007/s10665-014-9690-9} {\bibfield  {journal} {\bibinfo
  {journal} {J. Eng. Math.}\ }\textbf {\bibinfo {volume} {88}},\ \bibinfo
  {pages} {1} (\bibinfo {year} {2014})}\BibitemShut {NoStop}%
\bibitem [{\citenamefont {Ishikawa}\ \emph {et~al.}(2006)\citenamefont
  {Ishikawa}, \citenamefont {Simmonds},\ and\ \citenamefont {Pedley}}]{Pedley}%
  \BibitemOpen
  \bibfield  {author} {\bibinfo {author} {\bibfnamefont {T.}~\bibnamefont
  {Ishikawa}}, \bibinfo {author} {\bibfnamefont {M.~P.}\ \bibnamefont
  {Simmonds}}, \ and\ \bibinfo {author} {\bibfnamefont {T.~J.}\ \bibnamefont
  {Pedley}},\ }\href {\doibase 10.1017/S0022112006002631} {\bibfield  {journal}
  {\bibinfo  {journal} {J. Fluid Mech.}\ }\textbf {\bibinfo {volume} {568}},\
  \bibinfo {pages} {119} (\bibinfo {year} {2006})}\BibitemShut {NoStop}%
\bibitem [{\citenamefont {Llopis}\ and\ \citenamefont
  {Pagonabarraga}(2008)}]{Llopis_2008}%
  \BibitemOpen
  \bibfield  {author} {\bibinfo {author} {\bibfnamefont {I.}~\bibnamefont
  {Llopis}}\ and\ \bibinfo {author} {\bibfnamefont {I.}~\bibnamefont
  {Pagonabarraga}},\ }\href {\doibase 10.1140/epje/i2007-10295-y} {\bibfield
  {journal} {\bibinfo  {journal} {Eur. Phys. J. E}\ }\textbf {\bibinfo {volume}
  {26}},\ \bibinfo {pages} {103} (\bibinfo {year} {2008})}\BibitemShut
  {NoStop}%
\bibitem [{\citenamefont {Sharifi-Mood}\ \emph {et~al.}(2013)\citenamefont
  {Sharifi-Mood}, \citenamefont {Koplik},\ and\ \citenamefont
  {Maldarelli}}]{Nima_1}%
  \BibitemOpen
  \bibfield  {author} {\bibinfo {author} {\bibfnamefont {N.}~\bibnamefont
  {Sharifi-Mood}}, \bibinfo {author} {\bibfnamefont {J.}~\bibnamefont
  {Koplik}}, \ and\ \bibinfo {author} {\bibfnamefont {C.}~\bibnamefont
  {Maldarelli}},\ }\href {\doibase 10.1063/1.4772978} {\bibfield  {journal}
  {\bibinfo  {journal} {Phys. Fluids}\ }\textbf {\bibinfo {volume} {25}},\
  \bibinfo {pages} {012001} (\bibinfo {year} {2013})}\BibitemShut {NoStop}%
\bibitem [{\citenamefont {Michelin}\ and\ \citenamefont
  {Lauga}(2014)}]{Lauga_Peclet}%
  \BibitemOpen
  \bibfield  {author} {\bibinfo {author} {\bibfnamefont {S.}~\bibnamefont
  {Michelin}}\ and\ \bibinfo {author} {\bibfnamefont {E.}~\bibnamefont
  {Lauga}},\ }\href {\doibase 10.1017/jfm.2014.158} {\bibfield  {journal}
  {\bibinfo  {journal} {J. Fluid Mech.}\ }\textbf {\bibinfo {volume} {747}},\
  \bibinfo {pages} {572} (\bibinfo {year} {2014})}\BibitemShut {NoStop}%
\bibitem [{\citenamefont {Sharifi-Mood}\ \emph {et~al.}(2016)\citenamefont
  {Sharifi-Mood}, \citenamefont {Mozaffari},\ and\ \citenamefont
  {C{\'o}rdova-Figueroa}}]{Nima_3}%
  \BibitemOpen
  \bibfield  {author} {\bibinfo {author} {\bibfnamefont {N.}~\bibnamefont
  {Sharifi-Mood}}, \bibinfo {author} {\bibfnamefont {A.}~\bibnamefont
  {Mozaffari}}, \ and\ \bibinfo {author} {\bibfnamefont {U.}~\bibnamefont
  {C{\'o}rdova-Figueroa}},\ }\href {\doibase 10.1017/jfm.2016.317} {\bibfield
  {journal} {\bibinfo  {journal} {J. Fluid Mech.}\ }\textbf {\bibinfo {volume}
  {798}},\ \bibinfo {pages} {910} (\bibinfo {year} {2016})}\BibitemShut
  {NoStop}%
\bibitem [{\citenamefont {Happel}\ and\ \citenamefont
  {Brenner}(1983)}]{hapbren83}%
  \BibitemOpen
  \bibfield  {author} {\bibinfo {author} {\bibfnamefont {J.}~\bibnamefont
  {Happel}}\ and\ \bibinfo {author} {\bibfnamefont {H.}~\bibnamefont
  {Brenner}},\ }\href@noop {} {\emph {\bibinfo {title} {Low Reynolds Number
  Hydrodynamics with Special Applications to Particulate Media}}}\ (\bibinfo
  {publisher} {Martinus Nijhoff Publishers},\ \bibinfo {year}
  {1983})\BibitemShut {NoStop}%
\bibitem [{\citenamefont {Kim}\ and\ \citenamefont {Karrila}(2005)}]{Kim}%
  \BibitemOpen
  \bibfield  {author} {\bibinfo {author} {\bibfnamefont {S.}~\bibnamefont
  {Kim}}\ and\ \bibinfo {author} {\bibfnamefont {S.}~\bibnamefont {Karrila}},\
  }\href@noop {} {\emph {\bibinfo {title} {Microhydrodynamics: principles and
  selected applications}}}\ (\bibinfo  {publisher} {Dover Publications},\
  \bibinfo {address} {Mineola, NY, USA},\ \bibinfo {year} {2005})\BibitemShut
  {NoStop}%
\bibitem [{\citenamefont {Jeffery}(1915)}]{Jeffery}%
  \BibitemOpen
  \bibfield  {author} {\bibinfo {author} {\bibfnamefont {G.~B.}\ \bibnamefont
  {Jeffery}},\ }\href {\doibase 10.1112/plms/s2_14.1.327} {\bibfield  {journal}
  {\bibinfo  {journal} {Proc. London Math. Soc.}\ }\textbf {\bibinfo {volume}
  {s2-14}},\ \bibinfo {pages} {327} (\bibinfo {year} {1915})}\BibitemShut
  {NoStop}%
\bibitem [{\citenamefont {Spielman}(1970)}]{Spielman}%
  \BibitemOpen
  \bibfield  {author} {\bibinfo {author} {\bibfnamefont {L.~A.}\ \bibnamefont
  {Spielman}},\ }\href {\doibase 10.1016/0021-9797(70)90008-1} {\bibfield
  {journal} {\bibinfo  {journal} {J. Colloid Interface Sci.}\ }\textbf
  {\bibinfo {volume} {33}},\ \bibinfo {pages} {562} (\bibinfo {year}
  {1970})}\BibitemShut {NoStop}%
\bibitem [{\citenamefont {O'Neill}\ and\ \citenamefont
  {Majumdar}(1970)}]{Majumdar}%
  \BibitemOpen
  \bibfield  {author} {\bibinfo {author} {\bibfnamefont {M.~E.}\ \bibnamefont
  {O'Neill}}\ and\ \bibinfo {author} {\bibfnamefont {R.}~\bibnamefont
  {Majumdar}},\ }\href {\doibase 10.1007/BF01590641} {\bibfield  {journal}
  {\bibinfo  {journal} {Z. Angew. Math. Phys.}\ }\textbf {\bibinfo {volume}
  {21}},\ \bibinfo {pages} {164} (\bibinfo {year} {1970})}\BibitemShut
  {NoStop}%
\bibitem [{\citenamefont {Michelin}\ and\ \citenamefont
  {Lauga}(2015)}]{Michelin2}%
  \BibitemOpen
  \bibfield  {author} {\bibinfo {author} {\bibfnamefont {S.}~\bibnamefont
  {Michelin}}\ and\ \bibinfo {author} {\bibfnamefont {E.}~\bibnamefont
  {Lauga}},\ }\href {\doibase 10.1140/epje/i2015-15007-6} {\bibfield  {journal}
  {\bibinfo  {journal} {Eur. Phys. J. E}\ }\textbf {\bibinfo {volume} {38}},\
  \bibinfo {pages} {7} (\bibinfo {year} {2015})}\BibitemShut {NoStop}%
\bibitem [{\citenamefont {Subramanian}\ and\ \citenamefont
  {Balasubramaniam}(2005)}]{Subramanian}%
  \BibitemOpen
  \bibfield  {author} {\bibinfo {author} {\bibfnamefont {S.}~\bibnamefont
  {Subramanian}}\ and\ \bibinfo {author} {\bibfnamefont {R.}~\bibnamefont
  {Balasubramaniam}},\ }\href@noop {} {\emph {\bibinfo {title} {The Motion of
  Bubbles and Drops in Reduced Gravity}}}\ (\bibinfo  {publisher} {Cambridge
  University Press},\ \bibinfo {address} {New York, USA},\ \bibinfo {year}
  {2005})\BibitemShut {NoStop}%
\bibitem [{\citenamefont {Uspal}\ \emph {et~al.}(2016)\citenamefont {Uspal},
  \citenamefont {Popescu}, \citenamefont {Dietrich},\ and\ \citenamefont
  {Tasinkevych}}]{Uspal_2016}%
  \BibitemOpen
  \bibfield  {author} {\bibinfo {author} {\bibfnamefont {W.}~\bibnamefont
  {Uspal}}, \bibinfo {author} {\bibfnamefont {M.}~\bibnamefont {Popescu}},
  \bibinfo {author} {\bibfnamefont {S.}~\bibnamefont {Dietrich}}, \ and\
  \bibinfo {author} {\bibfnamefont {M.}~\bibnamefont {Tasinkevych}},\ }\href
  {\doibase 10.1103/PhysRevLett.117.048002} {\bibfield  {journal} {\bibinfo
  {journal} {Phys. Rev. Lett.}\ }\textbf {\bibinfo {volume} {117}},\ \bibinfo
  {pages} {048002} (\bibinfo {year} {2016})}\BibitemShut {NoStop}%
\bibitem [{\citenamefont {Morrison}(1970)}]{Morrison_1970}%
  \BibitemOpen
  \bibfield  {author} {\bibinfo {author} {\bibfnamefont {F.}~\bibnamefont
  {Morrison}},\ }\href {\doibase 10.1016/0021-9797(70)90171-2} {\bibfield
  {journal} {\bibinfo  {journal} {J. Colloid Interface Sci.}\ }\textbf
  {\bibinfo {volume} {34}},\ \bibinfo {pages} {210} (\bibinfo {year}
  {1970})}\BibitemShut {NoStop}%
\bibitem [{\citenamefont {Saintillan}\ and\ \citenamefont
  {Shelley}(2008)}]{Saintillan_2008}%
  \BibitemOpen
  \bibfield  {author} {\bibinfo {author} {\bibfnamefont {D.}~\bibnamefont
  {Saintillan}}\ and\ \bibinfo {author} {\bibfnamefont {M.}~\bibnamefont
  {Shelley}},\ }\href {\doibase 10.1063/1.3041776} {\bibfield  {journal}
  {\bibinfo  {journal} {Phys. Fluids}\ }\textbf {\bibinfo {volume} {20}},\
  \bibinfo {pages} {123304} (\bibinfo {year} {2008})}\BibitemShut {NoStop}%
\end{thebibliography}
%
\end{document}